\documentclass[a4paper,12pt]{article}
\pdfoutput=1
\pdfoutput=1
\usepackage{jheppub}
\usepackage{amsthm}
\usepackage{graphicx}
\usepackage{color}
\usepackage{amsmath}
\usepackage{graphicx,textcomp,float,gensymb,wrapfig, enumitem,comment,dsfont,subfigure,framed,slashed,appendix,wrapfig,wasysym}
\usepackage{comment}
\usepackage[export]{adjustbox}
\usepackage[utf8]{inputenc} 
\usepackage{multirow}

\newcommand{\be}{\begin{equation}}
\newcommand{\ee}{\end{equation}}
\def\bsp#1\esp{\begin{split}#1\end{split}}

\newcommand{\bea}{\begin{eqnarray}}  
\newcommand{\eea}{\end{eqnarray}}

\title{Gravity-mediated Dark Matter in Clockwork/Linear Dilaton Extra-Dimensions}
\author[a]{Miguel G. Folgado,}
\author[a]{Andrea Donini,}
\author[a]{Nuria Rius}
\affiliation[a]{Departamento de F\'isica Te\'orica and IFIC, Universidad de Valencia-CSIC,
C/ Catedr\'atico Jos\'e Beltr\'an, 2, E-46980 Paterna, Spain}

\emailAdd{donini@ific.uv.es}
\emailAdd{migarfol@ific.uv.es}
\emailAdd{nuria.rius@ific.uv.es}

\keywords{}

\abstract{
We study   for the first time the possibility that Dark Matter (represented by particles with spin $0,1/2$ or $1$)
interacts gravitationally with Standard Model particles in an extra-dimensional Clockwork/Linear Dilaton model.
We assume that both, the Dark Matter and the Standard Model, are localized in the IR-brane and only interact via gravitational mediators, 
namely the Kaluza-Klein (KK) graviton and the radion/KK-dilaton modes.
We analyse in detail the Dark Matter annihilation channel into Standard Model particles and into two on-shell Kaluza-Klein towers 
(either two KK-gravitons, or two radion/KK-dilatons, or one of each), finding that it is possible to obtain the observed 
relic abundance  via thermal freeze-out for Dark Matter masses in the range $m_{\rm DM} \in [1, 15]$ TeV for a 5-dimensional gravitational scale $M_5$
ranging from 5 to a few hundreds of TeV, 
even after taking into account the bounds from LHC Run II and irrespectively of the DM particle spin. 
}

\begin{document}
\hfill {\tt FTUV-19-1128.3781,  IFIC/19-56}

\maketitle
\flushbottom

\section{Introduction}
\label{sec:intro}

The Standard Model of Fundamental Interactions is in a wonderful shape, after the discovery of the Higgs boson in 2012 \cite{Aad:2012tfa}, 
and it may very well be that a huge energy desert above the TeV will be painstakingly explored till we could get in contact with even a single
new particle. However, a reasonable hope can alter this unappealing landscape: there it must be something more than the 
Standard Model out there, as the Standard Model is not able to explain what Dark Matter is.  The Nature of Dark Matter (DM) is, indeed,
one of the longest long-standing puzzles  to be explained in order to claim that we have a ``complete" picture of 
the Universe. On one side, both from astrophysical and cosmological data (see, e.g., Ref.~\cite{Bertone:2004pz} and refs. therein), 
rather clear indications regarding the existence of some kind of matter that gravitates but that does not interact with other particles by any other detectable mean can be gathered. On the other hand, no candidate to fill the r\^ole of DM has yet been observed
in high-energy experiments at colliders, nor is present in the Standard Model (SM) spectrum. 
Extensions of the Standard Model usually do include some DM candidate, a stable (or long-lived, with a lifetime as long 
as the age of the Universe) particle, with very small or none interaction with Standard Model particles and with particles of its own kind.
These states are usually supposed to be rather heavy and are called ``WIMP's", or ``weakly interacting massive particles".
Examples of these are the neutralino in supersymmetric extensions of the SM \cite{Dimopoulos:1981zb} or the lightest Kaluza-Klein particle in Universal Extra-Dimensions \cite{Appelquist:2000nn}. The typical range of masses for these particles was expected to be 
$m_{\rm DM} \in [100,1000]$ GeV. However, searches for these heavy particles at the LHC have pushed bounds on the masses of the candidates above the TeV scale, into the multi-TeV region. Moreover, experiments searching for DM particles through their interactions
with a fixed target, or ``Direct Detection" (DD) experiments (see, e.g., Ref.~\cite{Cushman:2013zza}) or through their annihilation into Standard Model particles, or 
``Indirect Detection" (ID) experiments (see, e.g., Ref.~\cite{Cirelli:2010xx}) have thoroughly explored the 
$m_{\rm DM} \in [100,1000]$ GeV region, pushing constraints on the interaction cross-section between DM and SM particles to very 
small values. In addition to this, both DD and ID experiments have a rather limited sensitivity above the TeV, as they have been mostly designed to look for ${\cal O} (100)$ GeV particles.  Other hypotheses have, however, been advanced:
DM particles could indeed be  ``feebly interacting massive particles" (FIMP's) \cite{Hall:2009bx}, ``strongly
interacting massive particles" (SIMP's) \cite{Hochberg:2014dra} or ``axion-like" very light particles (ALP's) \cite{Dias:2014osa}. 
All of these new proposals try to explore the possibility that DM is made of particles lighter than the expected WIMP range, 
a region where the exclusion bounds from DD and ID experiments are much weaker.   

If we take seriously the possibility that DM is made of ${\cal O} (1)$ TeV particles other options can be considered, though.
One interesting option is that the interaction between DM and SM particles be only gravitational. Being, however, the gravitational 
coupling enhanced by the existence of more than 3 spatial dimensions. Several extra-dimensional models have been proposed
in the last twenty years to explain a troublesome feature of the Standard Model, nicknamed as the ``Hierarchy Problem", i.e. the 
large hierarchy between the electro-weak scale, $\Lambda_{\rm EW} \sim 250$ GeV, and the Planck scale, $M_P \sim 10^{19}$ GeV.
In short, the mass of a scalar particle (the Higgs boson) should be sensitive (through loops) to the scale at which the 
Standard Model may be replaced by a more fundamental theory. If there is no new physics between the energy frontier reached
by the LHC and the Planck scale, then the mass of the Higgs boson should be as large as the latter. Being the experimentally measured
mass of the Higgs $m_{\rm H} = {\cal O} (\Lambda_{\rm EW})$, either the SM is not an effective theory and it is, after all, the 
ultimate theory (something not very convincing, as the SM does not explain Dark Matter, Dark Energy, Baryogenesis, 
the source of neutrino masses and, of course, gravity) or an incredible amount of fine-tuning between loop corrections stabilizes $m_{\rm H}$ at its
value. Extra-dimensional models solve the hierarchy problem by either replacing the Planck scale $M_{\rm P}$ with a fundamental 
gravitational scale $M_{\rm D}$ (being $D$ the number of dimensions) that could be as low as a few TeV (Large Extra-Dimensions models, or LED, see Refs.~\cite{Antoniadis:1990ew,Antoniadis:1997zg,ArkaniHamed:1998rs,Antoniadis:1998ig,ArkaniHamed:1998nn}), or by ``warping" the space-time such that the effective Planck scale $\Lambda$ felt by particles of the SM is indeed much smaller than the fundamental scale $M_{\rm D}$, similar to $M_{\rm P}$ (see Refs.~\cite{Randall:1999ee,Randall:1999vf}), or by a mixture of the two options (see Refs.~\cite{Giudice:2016yja,Giudice:2017fmj}).

The possibility that Dark Matter particles, whatever they be, may have an {\em enhanced} gravitational interaction with SM particles has been studied mainly in the context of warped extra-dimensions. The idea was first advanced in Refs.~\cite{Lee:2013bua,Lee:2014caa} and subsequently studied in 
Refs.~\cite{Han:2015cty,Rueter:2017nbk, Kumar:2019iqs,Rizzo:2018obe,Rizzo:2018joy,Carrillo-Monteverde:2018phy,Kraml:2017atm, Arun:2018yhg, Arun:2017zap}. The generic conclusion of these papers was
that when all the matter content is localized in the so-called TeV (or infrared brane), after taking into account current LHC bounds
 it was not possible to achieve the observed Dark Matter relic abundance in warped models for scalar DM particles (whereas this was not the
case for fermion and vector Dark Matter). However, an important
caveat was that these conclusions were drawn assuming the DM particle being {\it lighter} than the first Kaluza-Klein graviton mode. 
In this case, the only kinematically available channel to deplete the Dark Matter density in the Early Universe is the annihilation of two DM particles
into two SM particles through virtual KK-graviton exchange. However, in Ref.~\cite{Folgado:2019sgz}, we performed a 
check of the literature for the particular case of scalar DM in warped extra-dimensions, finding that as soon as the DM particle is allowed
to be {\it heavier} than the first KK-graviton, annihilation of two DM particles into two KK-gravitons becomes kinematically possible and, through
this channel, the observed relic abundance can indeed be achieved in a significant region of the parameter space within the freeze-out scenario. 
In the same paper, we included previously overlooked contributions to the DM annihilation cross-section, such as the possibility that DM annihilation 
into any pair of KK-gravitons can occur (regardless of the KK-number of the gravitons), and additional contributions to the thermally-averaged
cross-section arising at second order in the expansion of the metric around a background Minkowski 5-dimensional space-time (the correct order to 
reach, once considering production of two KK-gravitons). Eventually, we also study the impact of a Goldberger-Wise radion \cite{Goldberger:1999uk}, 
both in DM annihilation
through virtual radion exchange and through direct production of two radions. The region of the parameter space for which the observed DM 
relic abundance is achieved in the freeze-out framework corresponds to  DM masses in the range $m_{\rm DM} \in [1, 10]$ TeV, with first KK-graviton mass ranging from hundreds of GeV to some TeV.  The price to pay to achieve the freeze-out thermally-averaged cross-section is that the scale $\Lambda$ 
for which interactions between SM particles and KK-gravitons occur must be larger than 10 TeV, approximately. Therefore, in this scenario, the hierarchy
problem cannot be completeley solved and some hierarchy between $\Lambda$ and $\Lambda_{\rm EW}$ is still present. 
This is something, however, common to most proposals of new physics aiming at solving the hierarchy problem, 
as the LHC has found no hint whatsoever of new physics to date. One of the most interesting features of the scenario proposed in 
Ref.~\cite{Folgado:2019sgz} is that a large part of the allowed parameter space could 
be tested using either the LHC Run III or the HL-LHC data. By the end of the next decade, therefore, only tiny patches of the allowed 
parameter space should survive in case of no experimental signal, tipically corresponding to DM mass $m_{\rm DM} \sim 10$ TeV, 
near the theoretical unitarity bounds.

In this paper, we extend the study of DM in an extra-dimensional framework to the case of a 5-dimensional ClockWork/Linear Dilaton (CW/LD) model.
This model was proposed in Ref.~\cite{Giudice:2016yja} and its phenomenology at the LHC has been studied in Ref.~\cite{Giudice:2017fmj}. 
In this scenario, a KK-graviton tower with spacing very similar to that of LED models starts at a mass gap $k$ with respect to the zero-mode graviton. 
The fundamental gravitational scale $M_5$ can be as low as the TeV, where $k$ is typically chosen in the GeV to TeV range.
To our knowledge, this paper is the first attempt to use the CW/LD framework to explain the observed Dark Matter abundance in the Universe. 
In order to study this possibility, we very much follow the outline of our previous paper on DM in warped extra-dimensions
albeit  in this case we will consider DM particles with spin 0, $1/2$ and $1$. Also in this scenario we have found that
the freeze-out thermal relic abundance can be achieved in a significant region of the model parameter space, with the DM mass 
ranging from 1 TeV 
to approximately 15 TeV, for DM of any spin. The fundamental gravitational scale $M_5$
needed to achieve the target relic abundance goes from a few TeV to a few hundreds of TeV, thus introducing a little hierarchy problem.
Notice that the LHC Run III data and those of the high-luminosity upgrade HL-LHC will be able to test most of this region. 

The paper is organized as follows: in Sect.~\ref{sec:CW} we outline the theoretical framework, reminding shortly the basic ingredients of the 
ClockWork/Linear Dilaton extra-dimensional scenario and of how dark matter can be included within this hypothesis; 
in Sect.~\ref{sec:annihilres} we show our results for the annihilation cross-sections
of DM particles into SM particles, KK-gravitons and radion/KK-dilatons;  
in Sect.~\ref{sec:bounds} we review the present experimental bounds on the parameters of the model (the fundamental Planck scale $M_5$,
 the mass gap $k$ and the DM mass $m_{\rm DM}$) from the LHC
and from direct and indirect searches of Dark Matter, and recall the theoretical constraints 
(coming from unitarity violation and effective field theory consistency); 
in Sect.~\ref{sec:results} we explore the allowed parameter space such that 
the correct relic abundance is achieved for DM particles; 
and, eventually, in Sect.~\ref{sec:con} we conclude. 
In the Appendices we give some of the mathematical expressions used in the paper: 
in App.~\ref{app:feynman} we give the Feynman rules for the theory considered here; 
in App.~\ref{app:decay} we give the expressions for the decay amplitudes of the KK-graviton; 
in App.~\ref{app:kksum} we remind how the sum over KK-modes is carried on;
and, eventually, in App.~\ref{app:annihil} we give the formul\ae \, relative to the annihilation
cross-sections of Dark Matter particles into Standard Model particles, KK-gravitons and radion/KK-dilatons.
\section{Theoretical framework}
\label{sec:CW}

In this Section, we first review the freeze-out mechanism that could produce the observed DM relic abundance in the Universe. 
We then sketch the basic ingredients of the ClockWork/Linear Dilaton Extra-Dimensions scenario (CW/LD) needed to compute
the thermally-averaged DM annihilation cross-section. 

\subsection{The DM Relic Abundance in the Freeze-Out scenario}
\label{sec:relic}

The fact that a significant fraction of the Universe energy appears in the form of a non-baryonic ({\em i.e.} electromagnetically inert) matter
is the outcome of experimental data ranging from astrophysical to cosmological scales. This component of the 
Universe energy density is called {\em Dark Matter} and, in the cosmological ``standard model", the $\Lambda$CDM, 
it is usually assumed to be represented by stable (or long-lived) heavy particles ({\em i.e.} non-relativistic, or ``cold"). 
Within the thermal DM production scenario, 
DM particles were in thermal equilibrium with the rest of SM particles in the Early Universe. The DM density 
is governed by the Boltzmann equation \cite{Kolb:1990vq}:
\be
\frac{dn_{\rm DM}}{dt} = -3 H(T) \, n_{\rm DM} - \left\langle \sigma v \right\rangle \left [ n_{\rm DM}^2 - (n_{\rm DM}^{eq})^2 \right] \, , 
\label{boltzmann_equation}
\ee
with $T$ the temperature and $H(T)$ the Hubble parameter as a function of the temperature. 
The Boltzmann equation depends on a term proportional to the Hubble expansion rate at temperature $T$ and a term proportional to the thermally-averaged cross-section, $\left\langle \sigma v \right\rangle$. 
To obtain the correct population of DM particles within this scenario, 
the rate of decay and annihilation of DM particles should be such that, below a certain temperature $T_{\rm FO}$, the DM density $n_{\rm DM} (T)$ ``freezes out" 
and thermal fluctuations cannot any longer modify it. 
This occurs when $\left\langle \sigma v \right\rangle \times n_{\rm DM}$ falls below $H(T)$, DM decouples from the rest of particles and leaves an approximately constant number density in the co-moving frame, called relic abundance.
The experimental value of the relic abundance can be derived starting from the DM density in the $\Lambda$CDM model. 
From Ref.~\cite{Aghanim:2018eyx} we have $\Omega_{\rm CDM} h^2 = 0.1198 \pm 0.0012$, being $h$ the Hubble parameter. 
Solving eq.~(\ref{boltzmann_equation}), it can be found for the thermally-averaged cross-section at the freeze-out 
$\left\langle \sigma_{\rm FO} \, v \right\rangle \simeq 2.2 \times 10^{-26}$ cm$^3$/s \cite{Steigman:2012nb}.

It is very common to compute $\left\langle \sigma v \right\rangle$ in a given model in the so-called {\em velocity expansion} ({\em i.e.} assuming small relative velocity between the two DM particles). However, this approximation may fail in the neighbourhood of resonances. In the CW/LD model, the virtual graviton exchange cross-section is indeed the result of an infinite sum of KK-graviton modes. For this reason, we computed the  value of 
$\left\langle \sigma v \right\rangle$ using the exact expression from Ref.~\cite{Gondolo:1990dk}:
\be
\left\langle \sigma v \right\rangle =\frac{1}{8m_S^4 \, T \, K_2^2(x)} \int_{4m_S^2}^{\infty} ds (s - 4m_S^2)\, \sqrt{s} \, \sigma(s) \, K_1\left(\frac{\sqrt{s}}{T}\right)\, ,
\label{thermal_average}
\ee
being $K_1$ and $K_2$ the modified Bessel functions and $v$ the relative velocity between DM particles.

\subsection{A short summary on ClockWork/Linear Dilaton Extra-Dimensions}
\label{sec:clockwork}

The metric considered in the CW/LD scenario (see Refs.~\cite{Giudice:2016yja,Giudice:2017fmj}) is: 
\be
\label{eq:5dmetric}
ds^2 = e^{4/3 k r_c |y|} \left ( \eta_{\mu \nu} dx^{\mu}dx^{\nu} - r_c^2 \, dy^2 \right ) \, ,
\ee
where the signature of the metric is $(+,-,-,-,-)$ and, as usual, we use capital latin indices $M,N$ to run over the 5 dimensions and 
greek indices $\mu,\nu$ only over 4 dimensions. Notice that we have rescaled the coordinate in the extra-dimension such that $y$ is adimensional.
This particular metric was first proposed in the context of {\em Linear Dilaton} (LD) 
models and {\em Little String Theory} (see, {\em e.g.} Refs.~\cite{Antoniadis:2011qw,Baryakhtar:2012wj,Cox:2012ee} and references therein). 
The metric in eq.~(\ref{eq:5dmetric}) implies that the space-time is non-factorizable, as the length scales on our 4-dimensional space-time depending 
on the particular position in the extra-dimension due to the warping factor $\exp (2/3 \, k r_c \, |y|)$. Notice, however, that in the limit $k \to 0$ the standard, 
factorizable, flat LED case \cite{Antoniadis:1990ew,Antoniadis:1997zg,ArkaniHamed:1998rs,Antoniadis:1998ig,ArkaniHamed:1998nn} is immediately recovered. 
As for the case of the Randall-Sundrum model, also in the CW/LD scenario the extra-dimension is 
compactified on a ${\cal S}_1/{\cal Z}_2$ orbifold (with $r_c$ the compactification radius), and two branes
are located at the fixed points of the orbifold, $y = 0$ (``IR" brane) and at $y = \pi$ (``UV" brane).
Standard model fields are located in one of the two branes (usually the IR-brane).
The scale $k$, also called the ``clockwork spring" (a term inherited by its r\^ole in the discrete version of the Clockwork model \cite{Giudice:2016yja}), 
is the curvature along the 5th-dimension and it can be much smaller than the Planck scale (indeed, it can be as light as a few GeV). 
Being the relation between $M_{\rm P}$ and the fundamental gravitational scale $M_5$ in the CW/LD model: 
\be
M_{\rm P}^2 = \frac{M_5^3}{k} \left ( e^{2 \pi k r_c}  - 1 \right ) \, ,
\ee
it can be shown that, in order to solve or alleviate the hierarchy problem, $k$ and $r_c$ must satisfy the following relation:
\begin{equation}
k \, r_c = 10 + \frac{1}{2 \pi} \, \ln \left ( \frac{k}{\rm TeV} \right ) - \frac{3}{2 \pi} \, \ln \left ( \frac{M_5}{ 10 \, {\rm TeV}} \right ) \, .
\end{equation}
For $M_5 = 10$ TeV and $r_c$ saturating the present experimental bound on deviations from the Newton's law, $r_c \sim 100 \, \mu$m 
\cite{Adelberger:2009zz}, this relation implies that $k$ could be as small as $ k \sim 2$ eV, and KK-graviton modes would therefore 
be as light as the eV, also. This ``extreme" scenario does not differ much from the LED case, but for the important difference that the hierarchy
problem could be solved with just one extra-dimension (for LED models, in order to bring $M_5$ down to the TeV scale, an astronomical 
lenght $r_c$ is needed and, thus, viable hierarchy-solving LED models start with at least 2 extra-dimensions). In the phenomenological 
application of the CW/LD model in the literature, however, $k$ is typically chosen above the GeV-scale and, therefore, $r_c$ is accordingly
diminished so as to escape direct observation. Notice that, differently from the case of warped extra-dimensions, where scales 
are all of the order of the Planck scale ($M_5, k \sim M_{\rm P}$) or within a few orders of magnitude, in the CW/LD scenario, both the fundamental
gravitational scale $M_5$ and the mass gap $k$ are much nearer to the electro-weak scale $\Lambda_{\rm EW}$ than to the Planck scale, 
as in the LED model.

The action in 5D is: 
\begin{equation}
S = S_{\rm gravity} + S_{\rm IR} + S_{\rm UV}
\end{equation}
where the gravitational part is, in the Jordan frame:
\begin{equation}
\label{eq:5dgravity}
S_{\rm gravity} = \frac{M_5^3}{2} \int d^4 x \, \int_0^\pi r_c dy \sqrt{G^{(5)}} \, e^S \, \left [ R^{(5)}  + G_{(5)}^{MN} \partial_M S \partial_N S + 4 k^2  \right ] \, ,
\end{equation}
with $G_{MN}^{(5)}$ and $R^{(5)}$ the 5-dimensional metric and Ricci scalar, respectively, and $S$ the (dimensionless) dilaton field, $S = 2 k r_c |y|$. 
We consider for the two brane actions  the following expressions: 
\begin{equation}
S_{\rm IR} = \int d^4 x \, \sqrt{- g_{\rm IR}^{(4)}} \, e^S \left \{ - f_{\rm IR}^4 + {\cal L}_{\rm SM} + {\cal L}_{\rm DM}  \right \}
\end{equation}
and
\begin{equation}
S_{\rm UV} = \int d^4 x \, \sqrt{- g_{\rm UV}^{(4)}} \, e^S \left \{ - f_{\rm UV}^4 + \dots \right \} \, ,
\end{equation}
where $f_{\rm IR}, f_{\rm UV}$ are the brane tensions for the two branes and $g_{\rm IR,UV}^{(4)}  = - G^{(5)} / G^{(5)}_{55}$ 
is the determinant of the induced metric on the IR- and UV-brane, respectively. 
Throughout the paper, we consider all the SM and DM fields localized on the IR-brane, whereas on the UV-brane we could have
any other physics that is Planck-suppressed. We assume that DM particles only interact with the SM particles gravitationally by
considering only DM singlets under the SM gauge group. More complicated DM spectra with several particles will also not be studied here. 

Notice that the gravitational action is not in its canonical form. Going to the Einstein frame changing $G^{(5)}_{MN} \to \exp(-2/3 S) G^{(5)}_{MN}$, we get :
\begin{eqnarray}
\label{eq:5dgravity2}
S_{\rm gravity} &=& \int d^4 x \, \int_0^\pi r_c dy \sqrt{-G^{(5)}} \,  \left \{ \frac{M_5^3}{2} 
\left [ R^{(5)}  - \frac{1}{3} G_{(5)}^{MN} \partial_M S \partial_N S + 4 e^{- \frac{2}{3} S} k^2  \right ] \right \}
\nonumber \\
&+& \int d^4 x \, \int_0^\pi r_c dy \sqrt{-g^{(4)}} \, e^{-\frac{S}{3}} 
\left \{ \delta (y - y_0) \left [ - f^4_{\rm IR} +  {\cal L}_{\rm SM} + {\cal L}_{\rm DM} \right ] - \delta (y - \pi) f^4_{\rm UV} \right \} \, , \nonumber \\
&&
\end{eqnarray}
where now the gravitational action is the Einstein action and from the kinetic term of the dilaton field we can read out that the physical field must be rescaled as $\left ( M_5^{3/2}/\sqrt{3}\right ) \, S$. 
Eventually, it is important to stress that, in the Einstein frame, the brane action terms still have an exponential dependence $e^{-S/3}$ 
from the dilaton field. 
This action has a shift symmetry $S \to S + {\rm const}$ in the limit $k \to 0$, that makes a small value of $k$ with respect to $M_5$ ``technically
natural'' in the 't Hooft sense. Using the action above in the Einstein frame, it can be shown that the metric in eq.~(\ref{eq:5dmetric}) can be
recovered as a classical background if the brane tensions are chosen as: 
\begin{equation}
\label{eq:kfinetuning}
f^4_{\rm IR} = - f^4_{\rm UV} = - 4 k \, M_5^3 \, .
\end{equation}

Notice that, in a pure 4-dimensional scenario, the gravitational interactions would be enormously suppressed by powers of the Planck mass, 
while in an   extra-dimensional one the gravitational interaction is actually enhanced. 
Expanding the metric at first order around its static solution, we have: 
\be
\label{eq:metricexpansion}
G^{(5)}_{MN} =  e^{2/3  S} (\eta_{MN} + \frac{2}{M_5^{2/3}} h_{MN}) \, .
\ee
The 4-dimensional component of the 5-dimensional field $h_{MN}$ can be expanded in a Kaluza-Klein tower of 4-dimensional fields as follows:
\begin{equation}
h_{\mu\nu} (x,y) = \sum_{n = 0}^\infty \frac{1}{\sqrt{\pi r_c}} h^n_{\mu\nu} (x) \, \chi_n (y) \, .
\end{equation}
The $h^n_{\mu\nu} (x)$ fields are the KK-modes of the 4-dimensional graviton and the $\chi_{n}(y)$ factors are their wavefunctions. 
Notice that in the 4-dimensional decomposition of the 5-dimensional metric, two other fields are generally present: the graviphoton, $h_{\mu5}$, and the graviscalar $h_{55}$. The KK-tower of the graviscalar is absent from the low-energy spectrum, as they are eaten by the KK-tower of graviphotons to get a mass (due to the spontaneous breaking of translational invariance caused by the presence of one or more branes). 
These are, in turn, eaten by the KK-gravitons to get a mass (having, thus, five degrees of freedom). 
The surviving graviphoton zero-mode does not couple with the energy-momentum tensor in the weak gravitational field limit \cite{Giudice:1998ck},
whereas the graviscalar zero-mode will generically mix with the radion needed to stabilize the extra-dimension size.

The eigenfunctions $\chi_n (y)$ can be computed by solving the equation of motion in the extra-dimension of the fields: 
\begin{equation}
\left [ \partial_y^2 - k^2 r_c^2 + m_n^2 r_c^2 \right ] e^{k r_c |y|} \, \chi_n (y) = 0
\end{equation}
with Neumann boundary conditions $\partial_y \chi_n (y) = 0$ at $y = 0$ and $\pi$. Normalizing the eigenmodes such that the KK-modes have
canonical kinetic terms in 4-dimensions, we get: 
\begin{equation}
\left \{
\begin{array}{lll}
\chi_0 (y) & = & \sqrt{\frac{\pi k r_c}{e^{2 \pi k r_c} - 1}} \, , \\
&& \\
\chi_n (y) & = & \frac{n}{m_n r_c} e^{- k r_c |y|} \left ( \frac{k r_c}{n} \sin n |y| + \cos n |y| \right ) \, ,
\end{array}
\right .
\end{equation}
with masses
\begin{equation}
m_0^2 = 0 \, ; \qquad m_n^2 = k^2 + \frac{n^2}{r_c^2} \, .
\end{equation}

At the IR-brane one gets: 
\be
\mathcal{L} = -\frac{1}{M_5^{3/2}} T^{\mu \nu}(x) h_{\mu \nu}(x,y=0) = - \sum_{n=0} \frac{1}{\Lambda_n}  h^{n}_{\mu \nu} (x) \, T^{\mu \nu}(x)\, ,
\ee
where 
\begin{equation}
\left \{
\begin{array}{lll}
\frac{1}{\Lambda_0} &=& \frac{1}{M_{\rm P}} \, ,\\
&& \\
\frac{1}{\Lambda_n} &=&  \frac{1}{\sqrt{M_5^3 \pi r_c}}  \left ( 1 + \frac{k^2 r_c^2}{n^2} \right )^{-1/2} 
=  \frac{1}{\sqrt{M_5^3 \pi r_c}} \left ( 1 - \frac{k^2}{m_n^2} \right )^{1/2}  \, \qquad n \neq 0 \, ,
\end{array}
\right .
\label{Lambda_graviton}
\end{equation}
from which it is clear that the coupling between KK-graviton modes with $n \neq 0$ is suppressed by the effective scale $\Lambda_n$
and not by the Planck scale, differently from the LED case and similarly to the Randall-Sundrum one.

It is useful to remind here the explicit form of the energy-momentum tensor for a scalar, fermion and vector field:
\begin{equation}
\left \{
\begin{array}{lll}
T_{\mu \nu}^\Phi &=& (\partial_\mu \Phi)^\dagger (\partial_\nu \Phi) + (\partial_\nu \Phi)^\dagger (\partial_\mu \Phi) - \eta_{\mu \nu} \left\lbrace (\partial_\rho \Phi)^\dagger (\partial\rho \Phi) - m_\Phi^2 \Phi \Phi^\dagger \right\rbrace \, , \nonumber \\
&& \\
T_{\mu \nu}^\psi &=& 4 \left[ -\eta_{\mu \nu } \left\lbrace  \bar{\psi}(i\gamma_\rho\partial^\rho - m_\psi)\psi - \frac{1}{2} \partial^\rho (\bar{f}i\gamma_\nu f) \right\rbrace +  \left\lbrace \frac{1}{2}\bar{\psi} i \gamma_\mu\partial_\nu \psi - \frac{1}{4}\partial_\mu (\bar{\psi} i \gamma_\nu \psi) \right. \right. \, \nonumber \\
&& \\
&& + \left. \left. \frac{1}{2}\bar{\psi} i \gamma_\nu\partial_\mu \psi - \frac{1}{4}\partial_\nu (\bar{\psi} i \gamma_\mu \psi) \right\rbrace  \right], \, \nonumber \\
&& \\
T_{\mu \nu}^V &=& \left[ \eta_{\mu\nu} \left\lbrace \frac{1}{4} {\bf F}_{\rho \sigma} \, {\bf F}^{\rho \sigma} - \frac{m_V^2}{2}V^\rho V_\rho   \right\rbrace 
- {\bf F}_\mu^\rho \, { \bf F}_{\nu \rho} + m_V^2 V_\mu V_\nu \right] \, \\
\end{array}
\right .
\end{equation}
where
\begin{equation}
{\bf F}_{\mu\nu} = F_{\mu \nu} = \partial_\mu V_\nu - \partial_\nu V_\mu
\end{equation}
for an abelian gauge field and
\begin{equation}
{\bf F}_{\mu\nu} = F^a_{\mu \nu} = \partial_\mu V^a_\nu - \partial_\nu V^a_\mu + g f^{abc} V_\mu^b V_\nu^c
\end{equation}
for a non-abelian gauge field. In both cases, the expressions above refers to the unitary gauge. 
For the case of the SM massless gauge fields the expression is $T_{\mu \nu}^V\mid_{m_V = 0}$ (whilst we do not specify how the gauge field $V_\mu$ gets a mass).\\

\subsection{Introducing the radion}
\label{sec:rad}

Stabilization of the radius of the extra-dimension $r_c$ is an issue. In general (see, e.g., Refs.~\cite{Appelquist:1982zs,Appelquist:1983vs,deWit:1988ct}), bosonic quantum loops have a net effect on the boundaries of the extra-dimension such that the extra-dimension itself should shrink to a point. This feature, in a flat extra-dimension, can only be compensated by fermionic quantum loops and, usually, some supersymmetric framework is invoked to stabilize the radius of the extra-dimension 
(see, e.g., Ref.~\cite{Ponton:2001hq}). An additional advantage of supersymmetry in the bulk is that the CW/LD background metric may protect eq.~(\ref{eq:kfinetuning}) by fluctuations of the 5-dimensional cosmological constant (see, however, Ref.~\cite{Teresi:2018eai} for a non-supersymmetric clockwork implementation). 

In the CW/LD scenario we can use the already present bulk dilaton field $S$ to stabilize the compactification radius. 
If localized brane interactions generate a potential for $S$ at $y = \pi$, 
then we could fix the value of the field $S$ at the UV-brane, $ S_{\rm UV} = S\mid_\pi$. This is indeed an additional boundary condition that fixes
the distance between the two branes to be $\pi k \,  r_c = S_{\rm UV}/2$ \cite{Giudice:2016yja}:
\begin{equation}
\label{eq:localizedpotentials}
\left \{
\begin{array}{l}
S_{\rm IR} = \int d^4 x \, \sqrt{- g_{\rm IR}^{(4)}} \, e^S \left \{ - f_{\rm IR}^4 + \frac{\mu_{\rm IR}}{2} \left ( S - S_{\rm IR} \right )^2 + {\cal L}_{\rm SM} + {\cal L}_{\rm DM}  \right \} \, , \\
\\
S_{\rm UV} = \int d^4 x \, \sqrt{- g_{\rm UV}^{(4)}} \, e^S \left \{ - f_{\rm UV}^4 + \frac{\mu_{\rm UV}}{2} \left ( S - S_{\rm UV} \right )^2 + \dots \right \} \, ,
\end{array}
\right .
\end{equation}
with $\mu_{\rm IR}$ and $\mu_{\rm UV}$ two parameters with the dimension of a mass. In order to compute the scalar spectrum, we should introduce quantum fluctuations over the background values of $S (x,y) = S_0(y) + \varphi (x,y)$ (where $S_0 (y) = 2 k r_c |y|$) and of the metric, eq.~(\ref{eq:metricexpansion}). After deriving the Einstein equations for the two scalar degrees of freedom, $\varphi$ 
and\footnote{Using the notation of  Ref.~\cite{Cox:2012ee}, we call $\Phi$ the graviscalar $h_{55}$. Remember, however, that
after compactification the KK-tower of $h_{55}$ is eaten to give a longitudinal component to the KK-tower of gravitons.} $\Phi$, and imposing
the junction conditions at the boundaries, it can be shown that both satify the following equation of motion: 
\begin{equation}
\left [ \Box + \frac{1}{r_c^2} \frac{d^2}{dy^2} - k^2 \right ] e^{k r_c y} \left ( 
\begin{array}{l}
\Phi (x, y) \\
\varphi (x,y) 
\end{array}
\right ) = 0 \, .
\end{equation}
Notice that only the combination $v (x,y) = \sqrt{6} e^{k r_c y} M_5^{3/2} \left [ \Phi (x,y) - \varphi(x,y)/3 \right ]$ has a canonical kinetic term. 

Expanding $\Phi$ and $\varphi$ over a 4-dimensional plane-waves basis, 
\begin{equation}
\Phi (x,y) = \sum_n \Phi_n (y) Q_n (x) \, ; \qquad \varphi (x,y) = \sum_n \varphi_n (y) Q_n (x) \, ; \qquad \left [ \Box - m_{\Phi_n}^2 \right ] Q_n = 0 \, ,
\end{equation}
we can eventually derive the scalar fluctuations wave-functions (for example, in $\Phi$): 
\begin{equation}
\Phi_n (y) = N_n e^{- k r_c y} \left [ \sin (\beta_n y) + \omega_n \cos (\beta_n y) \right ]  \, ,
\end{equation}
with $N_n$ a normalization factor, $\beta_n = m_{\Phi_n}^2 - k^2$, and
\begin{equation}
\omega_n = - \frac{3 \beta_n \mu_T}{2 (k^2 + \beta_n^2) + k \mu_T} \, .
\end{equation}
In the so-called {\em rigid limit}, $\mu_{\rm UV} \to \infty$, the scalar spectrum is given by: 
\begin{equation}
\left \{ 
\begin{array}{l}
m_r^2 \equiv m_{\Phi_0}^2 = \frac{8}{9} k^2 \, ,\\
\\
m_{\Phi_n}^2 = k^2 + \frac{n^2}{r_c^2} \qquad (n \geq 1) \, , 
\end{array}
\right .
\end{equation}
first obtained in Ref.~\cite{Kofman:2004tk}, where we have identified the radion as the lightest state. 
Out of the rigid limit, the spectrum can be obtained expanding in inverse powers of $\mu_{\rm UV}$, 
introducing the adimensional parameters $\epsilon_{\rm IR, UV} = 2 k / \mu_{\rm IR, UV}$. At first order in the $\epsilon$'s, 
\begin{equation}
\left \{ 
\begin{array}{l}
m_r^2 \equiv m_{\Phi_0}^2 = \frac{8}{9} k^2 \left ( 1 - \frac{2 \epsilon_{\rm UV}}{9} \right ) + {\cal O} (\epsilon^2) \, ,\\
\\
m_{\Phi_n}^2 = k^2 + \frac{n^2}{r_c^2} \left [ 1 - \frac{6 (n^2 + k^2 r_c^2) (\epsilon_{\rm UV} + \epsilon_{\rm IR})}{9 n^2 \pi k r_c + \pi k^3 r_c^3} \right ] + {\cal O} (\epsilon^2) \, .
\end{array}
\right .
\end{equation}
There are no massless states for non-vanishing $\mu$'s ({\em i.e.}, when the extra-dimension is stabilized). In the unstabilized regime
(for $\mu_{\rm UV}, \mu_{\rm IR} \to 0$), the graviscalar and lowest-lying dilaton mode decouple and we expect two massless modes. 


The interactions of the radion and of the dilaton KK-tower with SM fields arises \cite{Cox:2012ee} from the term: 
\begin{equation}
\int d^4 x \sqrt{- g^{(4)}} \, e^{-S/3} \left [ {\cal L}_{\rm SM} + {\cal L}_{\rm DM}\right ] \, .
\end{equation}
The main difference between the CW/LD case and the Randall-Sundrum case is that in the former case a dilaton dependence $e^{-S/3}$ is still present in the brane term action 
going from the Jordan frame to the Einstein frame. On the other hand, the Randall-Sundrum action is already in the Einstein frame (its gravitational action is in the canonical form)
and the brane action term couples to gravity minimally, {\em i.e.} through the $\sqrt{-g^{(4)}}$ coefficient, only. 

Expanding the background metric and the dilaton field at first order in quantum fluctuations, we get (after KK-decomposition): 
\begin{eqnarray}
S_{\rm int} &=& - \frac{1}{2} \sum_n \Phi_n (0 ) \int d^4 x \sqrt{- g_0^{(4)}} \left [ g_0^{(4)} \right ]^{\mu\nu} \left [ T^{\rm SM}_{\mu\nu} + T^{\rm DM}_{\mu\nu} \right ] Q_n \nonumber \\
&-&  \frac{1}{3} \sum_n \varphi_n (0 ) \int d^4 x \sqrt{- g_0^{(4)}} \left [ {\cal L}_{\rm SM} + {\cal L}_{\rm DM}  \right ] Q_n \, .
\end{eqnarray}
Notice that the scalar fluctuations of metric AND dilaton couple with 4-dimensional fields through the usual energy-momentum trace and with a direct coupling with the 4-dimensional lagrangian. This is different from the case of the Randall-Sundrum model, where only the first kind of coupling is present, being the radion of purely gravitational origin (see, for example, Ref.~\cite{Goldberger:1999un}). In the CW/LD model, thus, there are two kinds of coupling between the radion and the KK-dilaton fields and the 4-dimensional fields sitting on the IR-brane.
Again, at first order in $\epsilon_{\rm UV, IR}$, we get: 
\begin{equation}
\left \{
\begin{array}{lll}
\frac{1}{\Lambda_\Phi^0} \equiv \frac{\Phi_0(0)}{2} &=& \frac{1}{6} \sqrt{\frac{k}{M_5^3} } \left ( 1 + \frac{4}{9} \epsilon_{\rm UV} \right ) + {\cal O}(\epsilon^2) \, , \\
&&\\
\frac{1}{\Lambda_\Phi^n} \equiv \frac{\Phi_n(0)}{2} &=& \frac{2 k r_c n}{\sqrt{3 \pi M_5^3 r_c} }  \left ( n^2 + k^2 r_c^2 \right )^{-1/2}  \left ( 9 n^2 + k^2 r_c^2 \right )^{-1/2} (1 - \epsilon_{\rm UV}) + {\cal O}(\epsilon^2) \\
&=& \frac{2 }{\sqrt{27 \pi M_5^3 r_c} } \, \frac{k}{m_{\Phi_n}} \sqrt{\frac{1 - \frac{k^2}{m_{\Phi_n}^2}}{1 - \frac{8}{9} \frac{k^2}{m_{\Phi_n}^2}}}
 (1 - \epsilon_{\rm UV}) + {\cal O}(\epsilon^2)
\end{array}
\right .
\label{Lambda_radion}
\end{equation}
and
\begin{equation}
\left \{
\begin{array}{l}
\frac{1}{\Lambda_\varphi^0} \equiv \frac{\varphi_0 (0)}{3} = \frac{2}{27} \sqrt{ \frac{k}{M_5^3} } \epsilon_{\rm UV} + {\cal O}(\epsilon^2) \, , \\
\\
\frac{1}{\Lambda_\varphi^n} \equiv \frac{\varphi_n (0)}{3} = \frac{n}{k \sqrt{3 \pi M_5^3 r^3_c} }  
\left [ \frac{ \left ( n^2 + k^2 r_c^2 \right ) }{ \left ( 9 n^2 + k^2 r_c^2 \right ) } \right ]^{1/2} \epsilon_{\rm UV} + {\cal O}(\epsilon^2) \\
\end{array}
\right .
\end{equation}
In the rigid limit ($\mu_{\rm UV, IR} \to \infty$) the coupling of dilaton modes with the SM lagrangian vanishes
($1/\Lambda_\varphi^0, 1/\Lambda_\varphi^n \to 0$). 
In the rest of the paper, we will work in this limit in order to get a sound insight of how the radion and dilaton KK-modes may affect
the generation of the freeze-out thermal abundance. A complete study of the impact of scalar perturbations to the DM phenomenology
would imply considering general values for $\epsilon_{\rm UV}$ and $\epsilon_{\rm IR}$ and it is beyond the scope of this paper. 

A further simplification that we are going to consider is the following: in the presence of a scalar field on the brane (such as the Higgs field), 
a non-minimal coupling of the scalar with the Ricci scalar is not forbidden by any symmetry. This may arise as a new term in the action: 
\begin{equation}
\label{eq:nonminimalHiggsmixing}
\Delta S_{\rm IR} = \int d^4 x \sqrt{- g^{(4)}} e^{\varphi/3} \xi R H^\dagger H \, .
\end{equation}
Such term induces an additional kinetic mixing between the graviscalar $\Phi_0$, the lowest-lying dilaton $\varphi_0$ and the Higgs and, therefore, 
additional couplings with the SM fields. We will neglect this non-minimal coupling in the rest of the paper, taking $\xi = 0$. 

Summarizing, in the rigid limit and in the absence of a mixing between the Higgs and the other scalar fields, the scalar perturbation interaction 
lagrangian with SM and DM particles at first order is: 
\begin{equation}
{\cal L}^{\rm SM}_v = \sum_{n=0}^\infty \frac{1}{\Lambda_\Phi^n}  \left [
T_{\rm SM}+ \frac{\alpha_{EM} \, C_{EM}}{8\pi} F_{\mu\nu} F^{\mu\nu} + \frac{\alpha_{S}C_{3}}{8 \pi} \sum_a F^a_{\mu\nu} F^{a\mu\nu}
\right ] \, v_n \, , 
\label{radion_lag}
\end{equation}
where $r = v_0$ is the radion field and $v_n$ for $n \geq 1$ is the dilaton KK-tower, and $T_{\rm SM}$ is the trace of the SM energy-momentum tensor. The coefficients of the coupling between scalar perturbations and massless gauge fields are given in App.~\ref{app:radFR}. Notice that massless gauge fields do not contribute to the trace of the energy-momentum tensor,  but they generate effective couplings from two different sources: quarks and $W$ bosons loops contribution and the trace anomaly \cite{Blum:2014jca}. 


\subsection{Contributions to $\left\langle \sigma v \right\rangle$ in the CW/LD scenario}
\label{sec:contribs}

We are not assuming any particular spin for the DM particle; our only assumptions are that there is just one particle responsible for the
whole DM relic abundance and that this particle interacts with the SM only gravitationally. Therefore, in the following we label 
such 
 particles generically by ${\rm DM}$'s. The total annihilation cross-section is:
\bea
\label{eq:sigmaDMtot}
\sigma_{\rm th} &=& 
\sum_{\rm SM} \sigma_{\rm ve}({\rm DM} \, {\rm DM}  \rightarrow {\rm SM} \, {\rm SM}) 
+ \sum_{n=1} \sum_{m=1} \sigma_{GG} ({\rm DM} \, {\rm DM} \rightarrow G_n \, G_m) \nonumber \\
&+& \sum_{n=0} \sum_{m=0}  \sigma_{\Phi\Phi} ({\rm DM} \, {\rm DM} \rightarrow \Phi_m \, \Phi_n) 
+ \sum_{n=1} \sum_{m=0} \sigma_{G\Phi} ({\rm DM} \, {\rm DM} \rightarrow G_n \, \Phi_m) \nonumber \\
\, ,
\eea
where in the first term, $\sigma_{\rm ve}$ (``${\rm ve}$" stands for ``virtual exchange"), we sum over all SM particles. 
The second term, $\sigma_{GG}$, corresponds to DM annihilation into KK-gravitons $G_n$. Notice that we do not consider 
DM annihilation into zero-mode gravitons $G_0$, as it is Planck-suppressed. 
The third term, $\sigma_{\Phi\Phi}$, corresponds to DM annihilation into radions and KK-dilaton modes.
Eventually, the fourth term, $\sigma_{G\Phi}$, is the production of one tower of KK-gravitons in association with a tower of radion/KK-dilatons
(a channel previously overlooked in the literature on the subject).
Notice that the KK-number is not conserved in the second, third and fourth term of eq.~(\ref{eq:sigmaDMtot})
due to the explicit breaking of momentum conservation in the 5th-dimension induced by the brane terms and, therefore, we must sum over
all values of $(m,n)$ as long as the condition $2 m_{\rm DM} \geq m_n + m_m$ (being $m_n$ the mass of the $n$-th KK-graviton or radion/KK-dilaton) is fulfilled.
If the DM mass $m_{\rm DM}$ is smaller than the mass of the first KK-graviton and of the radion, only the first channel is open.
Formul\ae \, for the DM annihilation into SM particles through virtual KK-graviton and radion/KK-dilaton exchange
are given in App.~\ref{app:annihil} in the small relative velocity approximation, expanding the centre-of-mass energy $s$ 
around $s \simeq 4 m_{\rm DM}^2$. Notice that, when computing the contribution of the radion/KK-dilaton exchange and KK-graviton exchange to the annihilation DM cross-section into SM particles,  it is of the uttermost importance to take into account properly the decay width of the radion/KK-dilaton and of the KK-gravitons. Formul\ae \, for the radion/KK-dilaton and KK-graviton decays\footnote{Recall that, due to the breaking of translational invariance in the extra-dimension, the KK-number is not conserved and heavy KK-graviton and KK-dilaton modes can also decay into lighter KK-modes when kinematically allowed.} are given in App.~\ref{app:decay}.


If the DM mass is larger than the radion or the first KK-graviton mass\footnote{Notice that, in the rigid limit, 
both the radion/KK-dilaton and KK-graviton masses only depend on the parameter $k$ and $r_c$ that are chosen to solve the hierarchy problem, differently from the RS scenario where the radion mass is an additional free parameter of the model.}, 
$m_{\rm DM} \leq (m_r, m_{G_1})$, the direct production of KK-graviton and/or radion/KK-dilaton towers becomes possible
and the other three channels of eq.~(\ref{eq:sigmaDMtot}) open.
The analytic expressions for $\sigma_{GG} ({\rm DM} \, {\rm DM} \rightarrow G_m \, G_n)$, $\sigma_{G\Phi} ({\rm DM} \, {\rm DM} \rightarrow G_m \, \Phi_n)$ and $\sigma_{\Phi\Phi} ({\rm DM} \, {\rm DM} \rightarrow \Phi_m \, \Phi_n)$ in the small relative velocity approximation 
are given in App.~\ref{app:annihil}. 

A DM singlet could have other interactions with the SM besides the gravitational one, through several so-called ``portals". Such scenarios have been extensively studied in the literature and are strongly constrained (see for instance \cite{Escudero:2016gzx,Casas:2017jjg}  for recent analyses), so we will neglect those couplings and focus only on the gravitational mediators that have not been previously considered.

\section{DM annihilation cross-section in CW/LD model}
\label{sec:annihilres}

In this section we study in detail the different contributions to the thermally-averaged DM annihilation cross-section, comparing 
the results for scalar, fermion and vector DM particles.

As we reminded in the previous section, for relatively low DM particles mass the first annihilation channel to open is the annihilation into SM particles through KK-graviton or radion/KK-dilaton exchange. Differently from the RS case (see Ref.~\cite{Folgado:2019sgz}), both the virtual KK-graviton and radion/KK-dilaton exchange cross-sections do not behave as the sum of relatively independent channels with well-separated peaks, one per KK-mode. For the typical values of $M_5$ and $k$ that may solve the hierarchy problem, in the CW/LD case a huge number of KK-modes must be coherently summed in $\sigma_{\rm ve} ({\rm DM} \, {\rm DM} \to {\rm SM} \, {\rm SM})$. 

In order to understand easily the difference between the cross-sections for scalar, fermion and vector DM particles, we remind in Tab.~\ref{tab:orbitals} the
dependence of the thermally-averaged annihilation cross-section $\langle \sigma v \rangle$ on the relative velocity $v$, from App.~\ref{app:annihil}. Recall that $v$ acts as 
a suppression factor and, therefore, the larger the power to which it appears, the smaller the cross-section.
\begin{table}[]
\begin{center}
\begin{tabular}{c|c|c|c|}
\cline{2-4}
\multicolumn{1}{l|}{}                                      & \multicolumn{1}{l|}{Scalar} & \multicolumn{1}{l|}{Fermion} & \multicolumn{1}{l|}{Vector} \\ \hline
\multicolumn{1}{|c|}{Graviton Virtual Exchange}                 & $v^4$ (d)                          & $v^2$ (p)                              & $v^0$ (s)                              \\ \hline
\multicolumn{1}{|c|}{Radion/Dilatons Virtual Exchange}     & $v^0$ (s)                          & $v^2$ (p)                              & $v^0$ (s)                              \\ \hline
\multicolumn{1}{|c|}{Annihilation into Gravitons}                 & $v^0$ (s)                          & $v^0$ (s)                              & $v^0$ (s)                              \\ \hline
\multicolumn{1}{|c|}{Annihilation into Radion/Dilatons}       & $v^0$ (s)                          & $v^2$ (p)                              & $v^0$ (s)                              \\ \hline
\multicolumn{1}{|c|}{Annihilation into Dilaton + Graviton}   & $v^0$ (s)                          & $v^0$ (s)                               & $v^0$ (s)                              \\ \hline
\end{tabular}
\end{center}
\caption{Velocity dependence of the different DM annihilation channels and the corresponding s-, p- or d-wave.}
\label{tab:orbitals}
\end{table}

\begin{figure}[htbp]
\centering
\includegraphics[width=160mm]{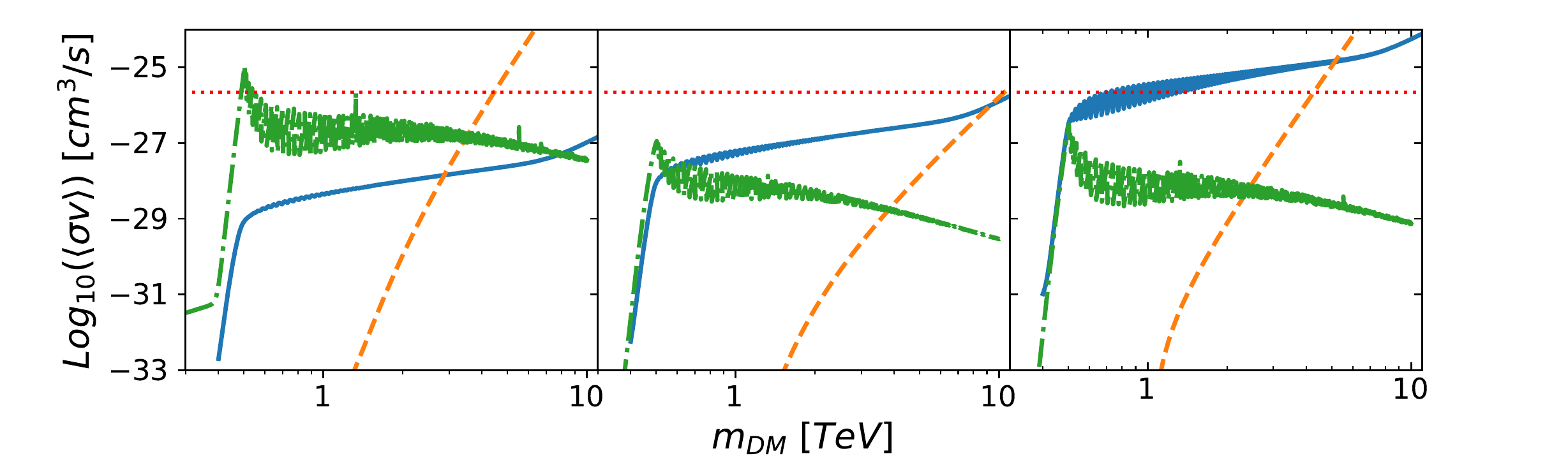}
\caption{\it Comparison of the thermally-averaged DM annihilation cross-section into SM particles through virtual radion/KK-dilaton exchange
$\left\langle \sigma_{{\rm ve},r} v \right\rangle$
(green dot-dashed lines) and virtual KK-graviton exchange $\left\langle \sigma_{{\rm ve},G} v \right\rangle$ (blue solid lines), as a function of the DM particle mass, $m_{DM}$. 
Left panel: scalar DM. Middle panel: fermion DM. Right panel: vector DM.
In all panels, the orange dashed line represents the thermally-averaged DM annihilation cross-section into KK-gravitons, $\left\langle \sigma_{GG} v \right\rangle$,
summing over all kinematically allowed KK-gravitons in the final state. 
The horizontal red-dotted line represents 
$\left\langle \sigma_{\rm FO} v \right\rangle$. The results have been obtained for $M_5 = 7$  TeV and $k = 1$ TeV.}
 \label{fig:contribuciones}
\end{figure}

The thermally-averaged virtual exchange cross-section, 
$\left\langle \sigma_{\rm ve} v \right\rangle = \left\langle (\sigma_{{\rm ve}, G} + \sigma_{{\rm ve}, \Phi}) v \right\rangle$,
is depicted in Fig.~\ref{fig:contribuciones} for a scalar (left panel), a fermion (middle panel) and
a vector (right panel) DM particle, respectively, for the particular choice $k = 1$ TeV and $M_5 = 7$ TeV
\footnote{Although the observed DM relic density can be obtained for lower values of  $(k,M_5)$, our choice is motivated by the fact that these are currently allowed by LHC data, as we will see in the next section.}.
Virtual radion/KK-dilaton exchange is shown with (green) dot-dashed lines, virtual KK-graviton exchange with (blue) solid lines. 
In all cases, $\sigma_{\rm ve} ({\rm DM} \, {\rm DM} \to {\rm SM} \, {\rm SM})$ is extremely small below $m_{\rm DM} \sim 500$ GeV, 
whilst rapidly increasing when $m_{\rm DM}$ approaches half the mass of the lightest mode (the radion). 
From that point onward, for larger and larger DM masses the cross-section starts to rapidly oscillate crossing threshold after threshold with 
new KK-modes entering the game. This behaviour can be clearly seen in the dot-dashed lines representing radion/KK-dilaton virtual exchange, 
where the difference between on-peak and off-peak cross-section can be as large as one order of magnitude. 
The sum over KK-dilaton modes does not increase the cross-section going to larger DM masses, as interferences from the near-continuum of modes collectively result in a slow decrease of $\sigma_{{\rm ve}, \Phi}$ going from $m_{\rm DM} \sim 1$ TeV to $m_{\rm DM} \sim 10$ TeV. 
The KK-graviton exchange cross-section shows a different behaviour: the difference between on- and off-peak is extremely small, and the sum over 
virtual KK-graviton modes gives a net (albeit slow) increase of the cross-section going to larger DM masses. 
These results are common to scalar, fermion and vector DM particles.

In the three panels, we also show the DM annihilation cross-section into real KK-gravitons, represented by an (orange) dashed line, 
and the freeze-out thermally-averaged cross-section $\left\langle \sigma_{\rm FO} v \right\rangle$, represented by the horizontal red-dotted line .
The DM annihilation cross-section into two real radion/KK-dilaton towers and into one KK-graviton and one radion/KK-dilaton tower 
are not shown, as both are much smaller and, therefore, irrelevant.
For a scalar or a vector DM particle  the real KK-graviton production cross-sections are very similar. This component of the total
cross-section takes over both the radion/KK-dilaton and KK-graviton virtual exchange and rapidly dominates the total cross-section 
for $m_{\rm DM}$ above a few TeVs. 
On the other hand,  the fermion DM real KK-graviton production cross-section is substantially smaller than those for scalar and vector DM
particles in the considered range of $m_{\rm DM}$ and its growth with $m_{\rm DM}$ is much slower (the corresponding cross-sections can be found in App. 
\ref{app:annihiG}).
We can see that, 
for the considered values of $M_5$ and $k$, the total fermion DM annihilation cross-section is dominated by virtual KK-graviton exchange
up to $m_{\rm DM} \sim 10$ TeV.

Comparing the results for different spin of the DM particle, we see that the scalar DM case is the only one where, for relatively low DM masses, the radion/KK-dilaton virtual exchange cross-section actually dominates over the KK-graviton virtual exchange one. The difference between the two contributions can be as large as two orders of magnitude for $m_{\rm DM}$ smaller than a few TeV, whereas the two  become comparable for $m_{\rm DM} \sim 10$ TeV (at a scale where, however, the real KK-graviton production has already become the dominant process). In this particular scenario, as it was the case for the RS model, the thermally-averaged virtual KK-graviton exchange cross-section is much lower than 
$\left\langle \sigma_{\rm FO} v \right\rangle$. 
On the other hand, the virtual radion/KK-dilaton exchange cross-section can actually reach the target value for $m^2_{\rm DM} \sim m^2_r/4$ 
({\em i.e.} $m^2_{\rm DM} = 2/9 k^2$ in the rigid limit).
For fermion and vector DM particles, this is not the case: the virtual radion/KK-dilaton exchange cross-section is of the same order or smaller than the virtual KK-graviton exchange cross-section\footnote{This is the combined effect of the different $v$-dependence according to the DM particle spin and of numerical factors.}. In summary, for the particular choice of $k$ and $M_5$ shown in Fig.~\ref{fig:contribuciones},
for a scalar DM particle  the target freeze-out value $\left\langle \sigma_{\rm FO} v \right\rangle$ is achievable either through
virtual radion/KK-dilaton exchange for low $m_{\rm DM}$ or via real KK-graviton production for $m_{\rm DM}$ a few TeV;
for a fermion DM particle $\left\langle \sigma_{\rm FO} v \right\rangle$ is not achieved for $m_{\rm DM} < 10$ TeV; and, 
for a vector DM particle, the target relic abundance is achieved through virtual KK-graviton exchange for $m_{\rm DM} \sim 1$ TeV
(as it was found in the RS scenario  \cite{Lee:2013bua,Rueter:2017nbk}).

\begin{figure}[htbp]
\centering
\includegraphics[width=160mm]{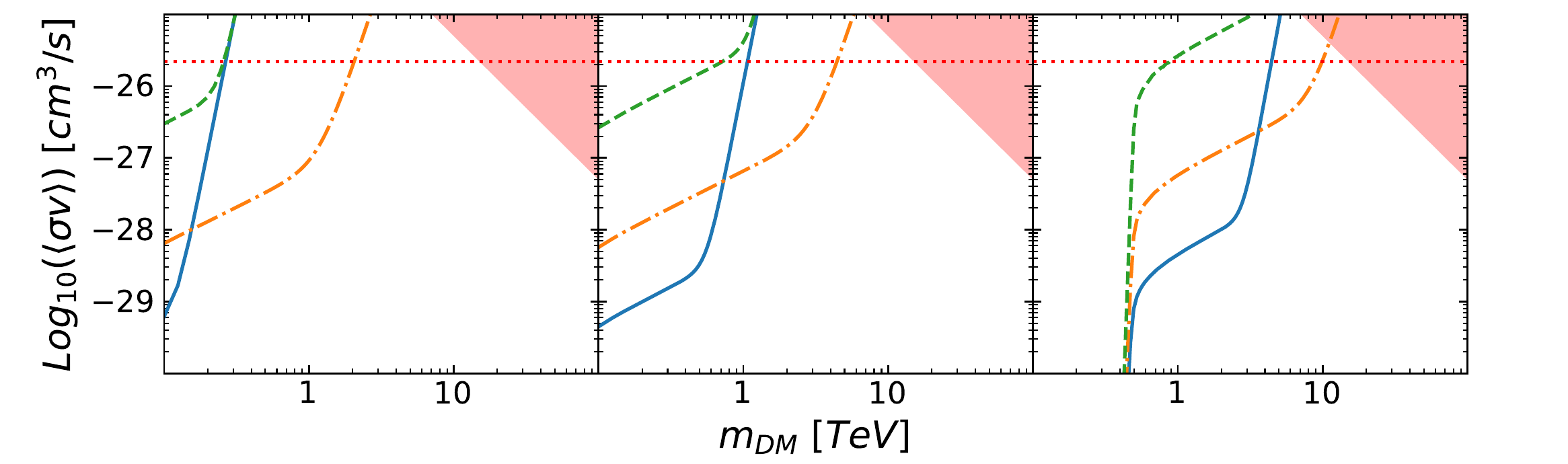}
\caption{\it The thermally-averaged DM annihilation cross-section through virtual
KK-graviton exchange and direct production of two KK-gravitons, $\sigma_G = \sigma_{{\rm ve},G} + \sigma_{G G}$, as a function of the DM 
mass $m_{DM}$ for three choices of $k$: $k = 10$ GeV (left panel); $k = 100$ GeV (middle panel); $k = 1000$ GeV (right panel).
In all panels, $M_5 = 7$ TeV.
The green dashed, orange dot-dashed and blue solid lines represent $\langle \sigma_G v \rangle$ for a vector, fermion and scalar DM particle, 
respectively. The red-shaded area represents the theoretical unitarity bound $\sigma \geq 1/s$. 
}
 \label{fig:conjunta}
\end{figure}

In Fig.~\ref{fig:conjunta} we show the total cross-section involving KK-gravitons, only (summing virtual KK-graviton exchange and KK-graviton production) as a function of the DM particle mass $m_{\rm DM}$ for different choices of $k$: 
$k = 10$ GeV (left panel), $k = 100$ GeV (middle panel) and $k = 1$ TeV (right panel). In all cases, $M_5 = 7$ TeV. 
In all panels, we plot $\langle \sigma_G v \rangle = \langle \left ( \sigma_{\rm ve, G} + \sigma_{GG} \right ) v \rangle$ for
scalar (blue, solid lines), fermionic (orange, dot-dashed lines) and vector (green, dashed lines) DM particles, thus making comparison easier. 
The red dotted horizontal line shows $\langle \sigma_{\rm FO} v \rangle$. 
For all choices of $k$,  at very low values of $m_{\rm DM}$ the scalar DM scenario give a much lower thermally-averaged
cross-section with respect to the fermion and vector case. It rapidly catches up, though, eventually merging with the vector case. We see that 
$\langle \sigma_G  v \rangle = \langle \sigma_{\rm FO} v \rangle$ at approximately $m_{\rm DM} \sim 10 \, k$
for $k$ below the TeV and $m_{\rm DM} = {\cal O} (k)$ for $k$ at the TeV in the scalar and vector case. 
On the other hand, a much larger value of $m_{\rm DM}$ is needed to achieve the freeze-out target value if the DM particle is a fermion.
The red-shaded area represents the theoretical unitarity
bound $\langle \sigma v \rangle \geq 1/s$, where we can no longer trust the theory outlined in Sect.~\ref{sec:CW} 
and higher-order operators should be taken into account. 

\begin{figure}[htbp]
\centering
\includegraphics[width=160mm]{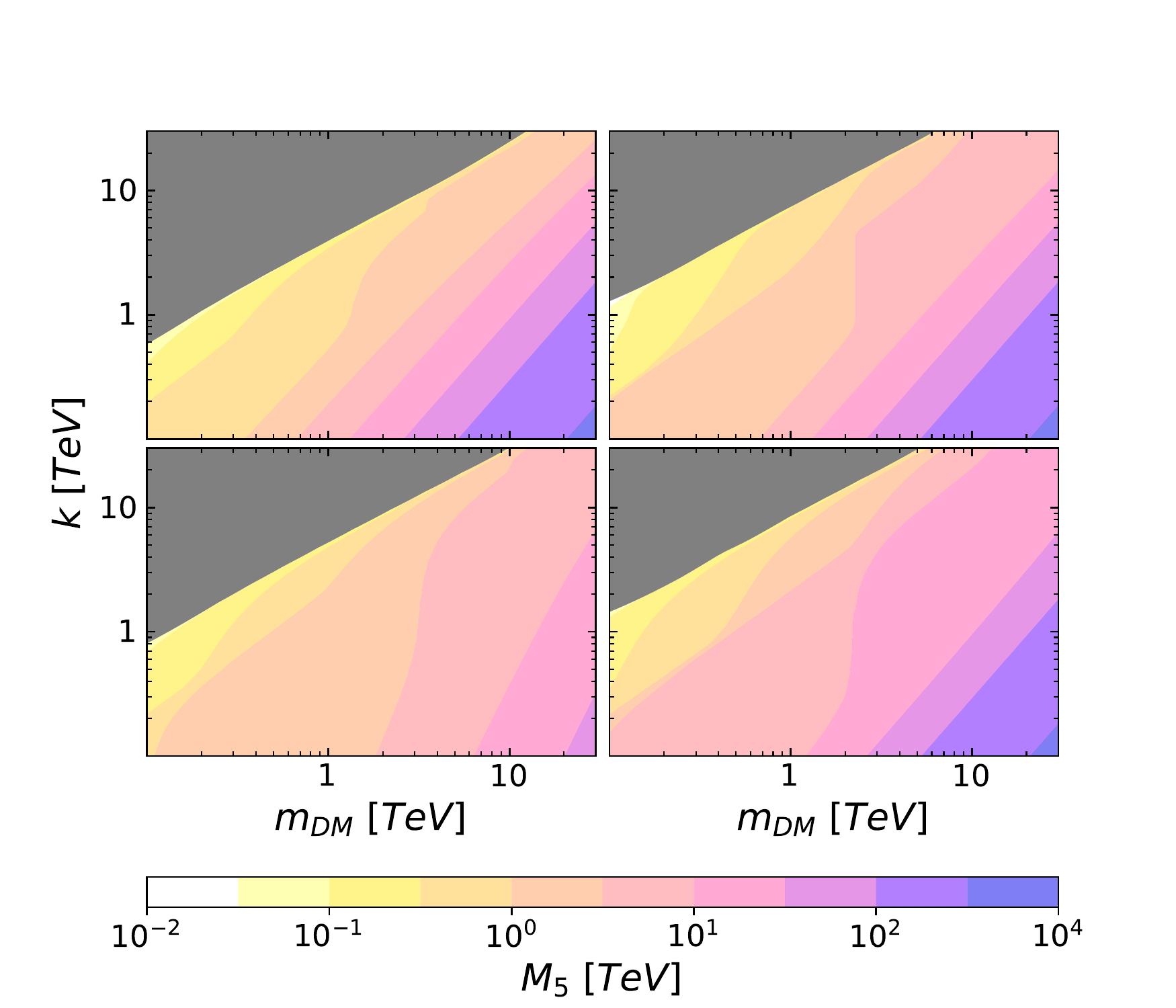}
\caption{\it Values of $M_5$ for which the correct DM relic abundance is obtained in the plane $m_{DM}, k$.
Top-left panel: Scalar DM particle, virtual KK-graviton exchange and real KK-graviton production only; 
Top-right panel: Scalar DM particle, virtual KK-graviton exchange and real KK-graviton production together with 
virtual radion/KK-dilaton exchange and real radion/KK-dilaton production;
Bottom-left panel: Fermion DM particle, virtual KK-graviton exchange and real KK-graviton production only; 
Bottom-right panel: Vector DM particle, virtual KK-graviton exchange and real KK-graviton production only.
The required $M_5$ ranges are shown by the color legend. The grey-shaded area represents the region of the 
parameter space for which is impossible to reach the freeze-out relic abundance.}
\label{fig:M5dependence}
\end{figure}

We have seen that it is relatively easy to achieve the freeze-out relic abundance for DM particles with a mass at the TeV scale or below for $M_5 = 7$ TeV. 
However, it is important to understand how this scales with $M_5$ so as to see how much having a DM candidate is compatible with solving
the hierarchy problem. This is shown in Fig.~\ref{fig:M5dependence}, where we draw the value of $M_5$ needed to achieve the
freeze-out DM annihilation cross-section $\langle \sigma_{\rm FO} v \rangle$ for a given choice of $k$ and $m_{\rm DM}$. 
In the top-left panel we show our results for a scalar DM particle using only virtual KK-graviton exchange and real KK-graviton 
production; in the top-right panel we again show our results for a scalar DM particle, albeit adding the contribution from virtual
radion/KK-dilaton exchange and real radion/KK-dilaton production (since we saw in Fig.~\ref{fig:contribuciones} that for this particular case
these contributions are quite relevant); in the bottom-left and bottom-right panels, on the
other hand, we show our results for a fermion and a vector DM particle, respectively, taking into account virtual KK-graviton exchange and real KK-graviton production only, as it was previously shown that in both cases the radion/KK-dilaton contribution is sub-dominant.
The grey area represents the region of the $(m_{\rm DM},k)$ plane for which it is not possible to achieve the freeze-out relic abundance. 
The coloured area is the region for which $\langle \sigma v \rangle$ can be as large as $\langle \sigma_{\rm FO} v \rangle$ for some
values of $m_{\rm DM}, k$ and $M_5$. The colour palette represents the corresponding ranges in $M_5$. 
The lowest values of $M_5$ for which we have $\langle \sigma v \rangle = \langle \sigma_{\rm FO} v \rangle$ are in the hundreds of GeV
range, whereas in the lower-right corner of all panels we find values of $M_5$ are of the order of tens of TeV.

\section{Experimental bounds and theoretical constraints}
\label{sec:bounds}

As we have seen in Fig.~\ref{fig:M5dependence}, the target relic abundance can be achieved in a 
vast region of the $(m_{\rm DM}, k)$ parameter space, if we allow $M_5$ to vary from $10^{-1}$ TeV to $10^2$ TeV.
However, experimental searches strongly constrain $k$ and $M_5$. 
We will summarize here the relevant experimental bounds and see how only a relatively small region of the parameter space 
is allowed, indeed.

\subsection{LHC bounds}
\label{sec:LHCbounds}

\begin{figure}[htbp]
\centering
\includegraphics[width=100mm]{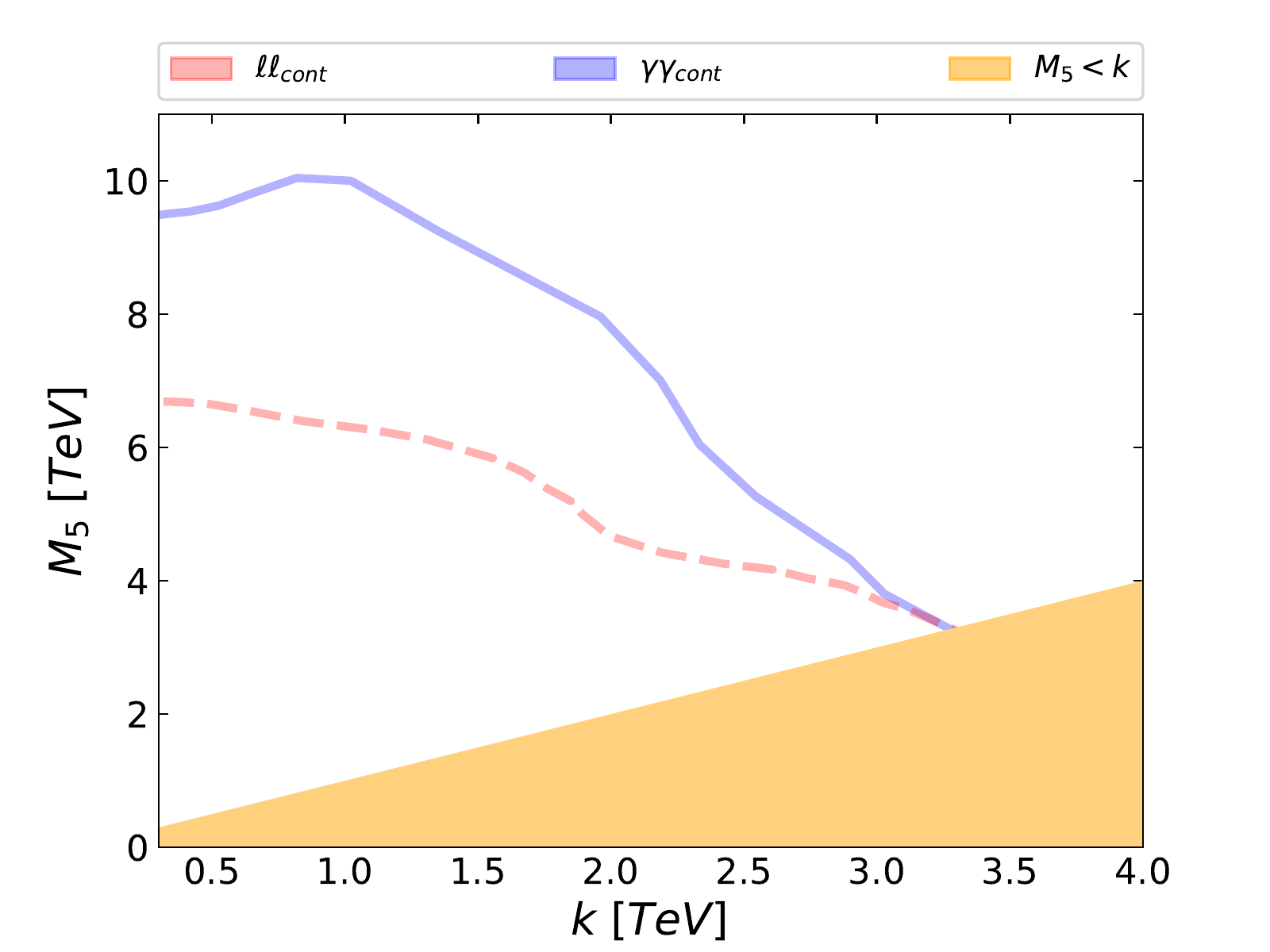}
\caption{\it Bounds in the $(k,M_5)$ plane from non-resonant searches at the LHC with $\sqrt{s} = 13$ TeV and 36 fb$^{-1}$,
from an analysis of ATLAS data \cite{Giudice:2017fmj} (dashed red line) and from the CMS Collaboration results \cite{CMS:2018thv} (solid blue line).
The orange-shaded area is the region of the parameter space for which $m_{G_1} \geq M_5$.}
\label{fig:LHCbounds}
\end{figure}

The strongest constraints are given by the non-resonant searches at LHC. Differently from the results from resonance searches at the LHC
\cite{Aaboud:2017yyg,ATLAS:2017wce}, data from non-resonant searches are not easily turned into bounds in $k$ and $M_5$.
We will therefore take advantage of the analysis performed in Ref.~\cite{Giudice:2017fmj} and of the dedicated analysis from the CMS 
Collaboration described in Ref.~\cite{CMS:2018thv}. The two bounds in the $(k,M_5)$ plane are shown in Fig.~\ref{fig:LHCbounds}, 
where the solid blue and dashed red lines represent results from  Ref.~\cite{Giudice:2017fmj} and Ref.~\cite{CMS:2018thv}, respectively.
The orange-shaded area is the region of the parameter space for which the mass of the first KK-graviton $m_{G_1}$ (where $m_{G_1} = k$) is larger than 
the scale of the theory, $M_5$. In this region of the parameter space the low-energy gravity effective theory is not trustable (see Sect.~\ref{sec:unitarityandOPE}).
In the rest of the paper, we have applied the experimental LHC bounds from Ref.~\cite{CMS:2018thv} as a conservative choice.

\subsection{Direct and Indirect Dark Matter Detection}
\label{sec:direct}

In order to understand the bounds from Direct Detection Dark Matter searches (DD) 
we need to compute the total cross-section for spin indepedent elastic scattering between Dark Matter and the nuclei \cite{Carrillo-Monteverde:2018phy}:
\be
\sigma_{{\rm DM}-p}^{\rm SI} = \left [ \frac{m_p \, m_{DM} }{A \pi (m_{DM} + m_p)} \right ]^2\left [ Af_p^{DM} + (A-Z)f_n^{S}  \right ]^2 \ , 
\label{eq:DD}
\ee
where $m_p$ is the proton mass, while  Z and A are the number of protons and the atomic number.
The nucleon form factors are given by the same formula for Dark Matter of any spin (at zero momentum transfer):
\be
\left \{
\begin{array}{lll}
f_p^{\rm DM} &=& \frac{m_{DM} \, m_p}{4 m_{G_1}^2\, \Lambda^2} 
\left \{ \sum_{q=u,c,d,b,s} 3 \left [ q(2) + \bar{q}(2) \right ] + \sum_{q=u,d,s} \frac{1}{3}  f_{Tq}^p \right \} \, ,\\
&& \\
f_n^{\rm DM} &=& \frac{m_{DM} \, m_p}{4 m_{G_1}^2 \, \Lambda^2} 
\left \{ \sum_{q=u,c,d,b,s} 3 \left [ q(2) + \bar{q}(2) \right ] + \sum_{q=u,d,s} \frac{1}{3} f_{Tq}^n \right \} \, ,
\end{array}
\right .
\ee
with $q(2)$  the second moment of the quark distribution function
\be
q(2) = \int_0^1 dx \ x \ f_q(x)
\ee
and $f_{Tq}^{N = p, n}$  the mass fraction of light quarks in a nucleon: $f^p_{Tu} = 0.023$, $f^p_{Td} = 0.032$ and $f^p_{Ts} = 0.020$ for a proton and $f^n_{Tu} = 0.017$, $f^n_{Td} = 0.041$ and $f^n_{Ts} = 0.020$ 
for a neutron \cite{Hisano:2010yh}. 

\begin{figure}[htbp]
\centering
\includegraphics[width=100mm]{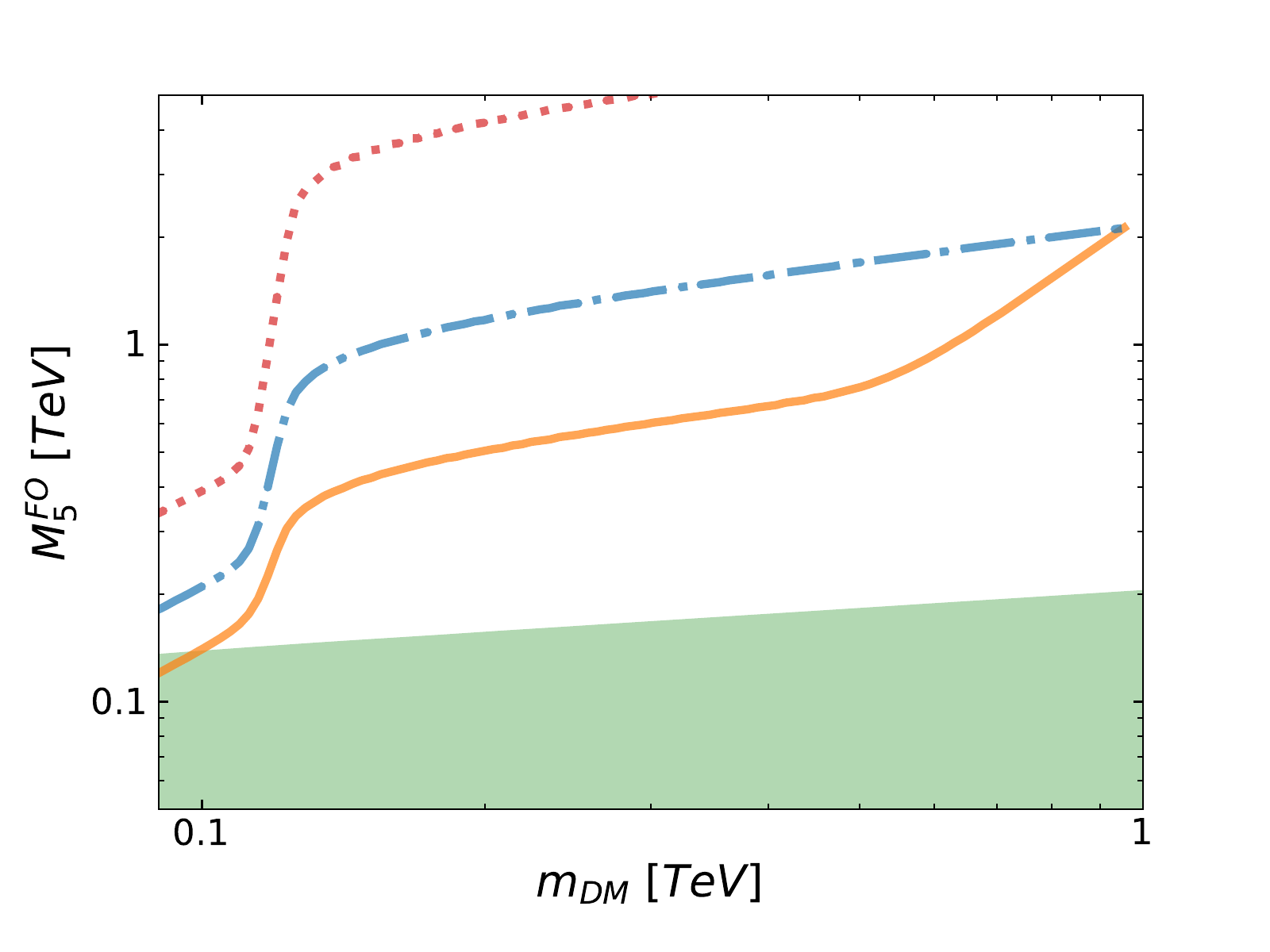}
\caption{The scale needed to achieve the freeze-out relic abundance, $M_5^{\rm FO}$, as a function of the DM mass $m_{\rm DM}$, for $k = 250$ GeV.
Solid orange, dot-dashed blue and dotted red lines correspond to scalar, fermion and vector DM, respectively. The green-shaded area, on the other hand, 
is the experimental bound in the ($m_{\rm DM}, M_5$) plane from XENON1T~\cite{Aprile:2017iyp}.
}
\label{fig:DDbounds}
\end{figure}

The strongest bounds come from the XENON1T experiment that uses $^{129}$Xe, ($Z=54$ and $A-Z=75$) as a target. 
In our analysis we compute the second moment of the PDF's using Ref.~\cite{Martin:2009iq} and the exclusion curve of 
XENON1T~\cite{Aprile:2017iyp} to set constraints in the parameter space. 
In Fig.~\ref{fig:DDbounds} we show the scale needed to achieve the freeze-out relic abundance, $M_5^{\rm FO}$, as a function of the DM mass $m_{\rm DM}$, for $k = 250$ GeV.
The three lines (solid orange, dot-dashed blue and dotted red) correspond to scalar, fermion and vector DM, respectively. The green-shaded area
is the experimental bound in the ($m_{\rm DM}, M_5$) plane from XENON1T. We can see that the bounds imposed by DD only constrain very low values of $m_{\rm DM}$ and 
they are irrelevant in the range of DM masses considered in the rest of this paper ($m_{\rm DM} \geq 100$ GeV). We have checked that this result is general also for 
other values of $k$.

With respect to Indirect Detection Dark Matter searches (ID), several experiments are analysing differents signals. For instance, the Fermi-LAT Collaboration studied 
the $\gamma$-ray flux arriving at Earth from the galactic center \cite{TheFermi-LAT:2015kwa,TheFermi-LAT:2017vmf} 
and from different Dwarf Spheroidal galaxies \cite{Fermi-LAT:2016uux}. 
Other experiments detect charged particles instead of photons, as it is the case of AMS-02 that presented data about the positron \cite{PhysRevLett.113.221102} 
and anti-proton fluxes coming from the galactic center \cite{PhysRevLett.117.091103}. These results are relevant in various DM models that can generate 
a continuum spectra of SM particles, such as our case. However, current data from ID only allows to constrain DM masses below 100 GeV, 
a region which is already excluded by LHC data.

\subsection{Theoretical constraints}
\label{sec:unitarityandOPE}

Besides the experimental limits, there are mainly two theoretical concerns about the validity of our calculations which affect part of the 
$(m_{\rm DM}, k,M_5)$ parameter space. The first one is related to the fact that we are performing just a tree-level computation of the relevant DM annihilation cross-sections, 
and we should worry about unitarity issues. In particular, the annihilation cross-section into a pair of real
KK-gravitons, $\sigma({\rm DM} \, {\rm DM} \to G_n G_m)$, diverges as $m_{\rm DM}^{10}/(m_{G_n}^4 m_{G_m}^4)$ for scalar and vector DM
and as $m_{\rm DM}^6/(m_{G_n}^2 m_{G_m}^2)$ for fermion DM (see eqs.~(\ref{graviton_real_scalar_dm},\ref{graviton_real_fermion_dm}) and (\ref{graviton_real_vector_dm}) 
in App.~\ref{app:annihiG}). When the DM mass becomes very large with respect to the KK-graviton masses, it is important to check that the effective theory is still unitary
\cite{Kahlhoefer:2015bea}. Asking for the cross-section to be bounded, $ \sigma < 1/s \simeq 1/m_{\rm DM}^2$, we got the red-shaded areas shown 
in Fig.~\ref{fig:conjunta}. If we combine the unitarity requirement with the request that the freeze-out thermally-averaged cross-section is achieved to get 
the correct DM relic abundance, we have an upper bound on the DM mass: $m_{\rm DM} \lesssim 1/\sqrt{\sigma_{\rm FO}}$, independently on the parameters that
determine the geometry of the space-time, ($k$ and $M_5$). This will be shown by a vertical line in the ($m_{\rm DM}, k$) plane in Fig.~\ref{fig:finalresults}.

The second theoretical issue refers to the consistency of the effective theory framework: in the CW/LD scenario, at energies somewhat 
larger than $M_5$ the KK-gravitons are strongly coupled and the five-dimensional field theory from which we start is no longer valid. We therefore impose that at least $m_{G_1} = k < M_5$ to trust  our results. Notice that this constraint is general for any effective field theory: since we are including the KK-graviton tower
in the low-energy spectrum, for the effective theory to make sense the cut-off scale $M_5$  should be larger  than the masses of such states.
For the same reason, we also ask for the Dark Matter mass $m_{\rm DM}$ to be lighter than $M_5$, $m_{\rm DM} < M_5$, although we will see that, 
in the allowed region, this requirement is almost always fulfilled.

\section{Results}
\label{sec:results}

\begin{figure}[htbp]
\centering
\includegraphics[width=160mm]{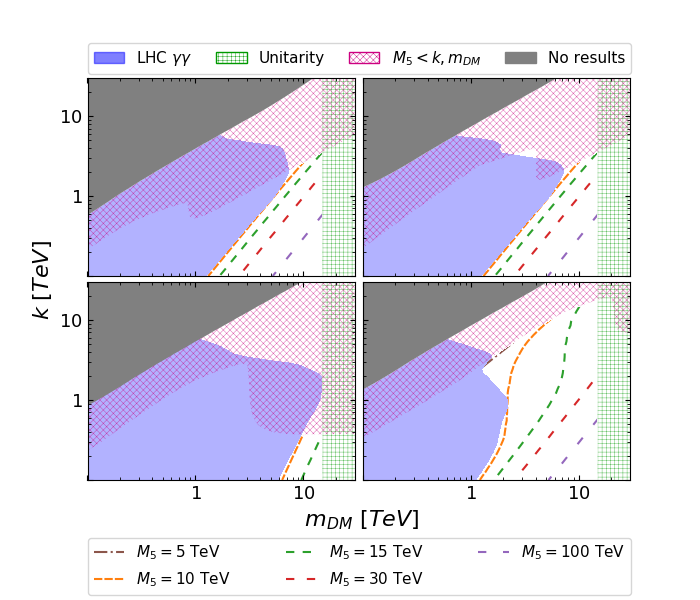}
\caption{\it Region of the $(m_{\rm DM}, k)$ plane for which $\left\langle \sigma v \right\rangle = \left\langle \sigma_{\rm FO} v \right\rangle$. 
Upper left panel: scalar DM (unstabilized extra-dimension);
Upper right panel:  scalar DM (stabilized extra-dimension in the rigid limit, $\epsilon_{\rm IR} = \epsilon_{\rm UV} = 0$, without
non-minimal coupling with gravity, $\xi =0$); 
Lower left panel: fermion DM (stabilized extra-dimension in the rigid limit without non-minimal coupling with gravity);
Lower right panel: vector DM (stabilized extra-dimension in the rigid limit without non-minimal coupling with gravity).
In all panels, the grey-shaded area represents the part of the parameter space for which it is impossible to achieve the correct relic abundance; 
the red diagonally-meshed area is the region for which the low-energy CW/LD effective theory is untrustable, as $M_5 < k$; 
the blue-shaded area is excluded by non-resonant searches at the LHC with 36 fb$^{-1}$ at $\sqrt{s} = 13$ TeV \cite{Giudice:2017fmj}; 
eventually, the green vertically-meshed area on the right is the region where the theoretical unitarity constraints are not fulfilled, 
$m_{\rm DM} \gtrsim 1/\sqrt{\sigma_{\rm FO}}$.
In all panels, the white area represents the region of the parameter space for which the correct relic abundance is achieved
(either through direct KK-graviton and/or radion/KK-dilaton production, as in the case of scalar DM, or through virtual KK-graviton exchange,
as for fermion and vector DM) and not excluded by experimental bounds and theoretical constraints.
The dashed lines depicted in the white region represent the values of $M_5$ needed to obtain the correct relic abundance
(from Fig.~\ref{fig:M5dependence}).
}
 \label{fig:finalresults}
\end{figure}

We show in Fig.~\ref{fig:finalresults} the allowed parameter space in the $(m_{\rm DM},k)$ plane
for which the target value of $\left\langle \sigma v \right\rangle$ needed to achieve the correct DM relic abundance 
in the freeze-out scenario, ($\left\langle \sigma_{\rm FO} v \right\rangle = 2.2 \times 10^{-26}$ cm$^3$/s), can be obtained, 
taking into account both the experimental bounds and the theoretical constraints outlined in Sec.~\ref{sec:bounds}.

In the upper left panel we show our results for a scalar DM particle, considering only decays into SM particles through virtual KK-graviton exchange
or into KK-gravitons. This corresponds to the unstabilized regime, {\em i.e.} when the coefficients $\mu_{\rm IR}, \mu_{\rm UV}$ 
of the localized potential terms in eq.~(\ref{eq:localizedpotentials}) vanish. In the upper right panel we show our results for scalar DM
when the extra-dimension is stabilized 
in the rigid limit, $\mu_{\rm IR}, \mu_{\rm UV} \to \infty$, and in the absence
of non-minimal coupling with gravity, $\xi = 0$ (see Sect.~\ref{sec:CW} for details). In this case, the annihilation of DM particles occurs
through virtual KK-graviton and radion/KK-dilaton exchange into SM particles and through direct KK-graviton and radion/KK-dilaton production. In the bottom left and right panels we show our results for a fermion and a vector DM particle, respectively. In both cases, 
the radion/KK-dilaton contribution (in the rigid limit with $\xi = 0$) is included but it is irrelevant.

As a guidance, dashed lines taken from Fig.~\ref{fig:M5dependence} represent the values of $M_5$ needed to achieve the relic abundance in a particular point of the $(m_{\rm DM}, k)$ plane. The legend for the four plots is given in the Figure caption.

\subsection{Scalar Dark Matter}
\label{sec:scalar DM}

In the case of scalar DM, depicted in the upper left and right panels, virtual KK-graviton exchange is not enough to 
achieve the freeze-out relic abundance. For this reason, when the extra-dimension is unstabilized (left panel),
$\left\langle \sigma_{\rm FO} v \right\rangle$ can be obtained only when the KK-graviton production channel opens, as it was
the case for the RS scenario \cite{Folgado:2019sgz}. As a consequence, the DM particle mass has to be in a given relation with the
mass of the KK-graviton tower and, therefore, a grey region for which it is impossible to achieve $\left\langle \sigma_{\rm FO} v \right\rangle$
can be seen. The red diagonally-meshed area represents the region of the parameter space for which the correct relic abundance is achieved
with a value of $M_5$ lower than the mass of the first KK-graviton, $m_{G_1} = k$. Above this line the low-energy effective theory 
we are using is untrustable, as new dynamical particles in the spectrum are heavier than the scale of the theory.
The blue-shaded area represents the excluded region from searches of non-resonant channels at LHC Run II 
with 36 pb$^-1$ from Ref.~\cite{Giudice:2017fmj}. 
The green vertically-meshed area is the upper bound on the DM mass that must be fulfilled to comply with unitarity.

When the extra-dimension is stabilized (right panel), the virtual radion/KK-dilaton exchange channel may reach the target value for the cross-section 
for some values of the DM mass for which the KK-graviton exchange channel may not (see Fig.~\ref{fig:contribuciones}). Therefore, a grey area is present
but it somewhat smaller than in the unstabilized case (differently from the Randall-Sundrum case, where no grey area was found in this case \cite{Folgado:2019sgz}). 
Most of this region is excluded because the value of $M_5$ is lower than $k$ and, thus, the effective theory we are using is untrustable (red-meshed region). 
As a consequence,  the allowed region that complies with experimental bounds and theoretical constraints is very similar to the unstabilized case
and, roughly speaking, corresponds to $m_{\rm DM} \in [1,15]$ TeV and $k < 6$ TeV. Within the allowed region, $M_5$ may vary between 10 TeV's and a few hundreds of TeV's.

\subsection{Fermion Dark Matter}
\label{sec:scalar DM}

The case of fermion DM is depicted in the lower left panel. The meaning of the coloured areas is the same as for the upper panels: 
the grey area is the region of the parameter space for which is impossible to achieve $\left\langle \sigma_{\rm FO} v \right\rangle$;
the blue-shaded area corresponds to the LHC Run II exclusion bound \cite{Giudice:2017fmj}; 
the red diagonally-meshed and green vertically-meshed areas represent theoretical unitarity bounds; 
and, the white area is the allowed region of the parameter space, where dashed lines represent benchmark
values of $M_5$ useful to understand its scaling. The main difference with the scalar (and vector) DM case is that for fermion DM 
a rather small region of the parameter space is compatible with all bounds and constraints. This is a consequence of the slower
dependence of the direct KK-graviton production cross-section with $m_{\rm DM}$ (see Figs.~\ref{fig:contribuciones} and \ref{fig:conjunta}
and eq.~(\ref{eq:graviton_real_fermion_dm_cruzados}) in App.~\ref{app:annihil}). Eventually, the allowed region that complies with experimental bounds and theoretical constraints corresponds to $m_{\rm DM} \in [4,15]$ TeV and $k < 1$ TeV. Within the allowed region, $M_5$ may vary between 10 TeV's and a few tens of TeV's.

\subsection{Vector Dark Matter}
\label{sec:scalar DM}

The case of vector DM is depicted in the lower right panel. The meaning of the coloured areas is the same as for the upper panels: 
the grey area is the region of the parameter space for which is impossible to achieve $\left\langle \sigma_{\rm FO} v \right\rangle$;
the blue-shaded area corresponds to the LHC Run II exclusion bound \cite{Giudice:2017fmj}; 
the red diagonally-meshed and green vertically-meshed areas represent theoretical unitarity bounds; 
and, the white area is the allowed region of the parameter space, where dashed lines represent benchmark
values of $M_5$ useful to understand its scaling. The main difference with the scalar and fermion DM case is that for vector DM 
it is possible to achieve the correct relic abundance through the virtual KK-graviton exchange channel,  and the requirements on $M_5$ are less stringent. 
As a consequence, a rather large region of the parameter space 
is compatible with all bounds and constraints. The allowed region that complies with experimental bounds and theoretical constraints corresponds to $m_{\rm DM} \in [0.6,15]$ TeV and $k$ may be as large as $\sim 20$ TeV. Within the allowed region, $M_5$ may vary between a 
5 TeV's and a few hundreds of TeV's.

\section{Conclusions}
\label{sec:con}

In this paper we have explored the possibility that the observed Dark Matter component in the Universe is represented by some new particle with mass in the TeV range
which interacts with the SM particles only gravitationally, in agreement with non-observation of DM signals at both direct and indirect detection DM experiments. 
In standard 4-dimensional gravity, the interaction between such DM particles and SM particles would be too feeble to reproduce the observed DM relic abundance. However, we have found that this is not the case once this setup is embedded in a Clockwork/Linear Dilaton scenario, along the ideas of the CW/LD proposal of Refs.~\cite{Giudice:2016yja,Giudice:2017fmj}. 
We consider two 4-dimensional branes in a 5-dimensional space-time with non-factorizable CW/LD metric \cite{Antoniadis:2011qw}
at a separation $r_c$, very small compared with present bounds on deviations from Newton's law. 
On one of the branes, the so-called ``IR-brane", both the SM particles and a DM particle (with spin $0, 1/2$ or $1$)
are confined, with no particle allowed to escape from the branes to explore the bulk. It can be shown that gravitational interaction
between particles on the IR-brane (in our case between a DM particle and any of the SM particles) 
occurs with an amplitude proportional to $1/M^2_{\rm P}$ when the two particles exchange a graviton zero-mode, 
but with a suppression factor $1/\Lambda_n^2$ when they interact exchanging the $n$-th KK-graviton mode.
As the effective coupling $\Lambda_n$ can be as low as a few TeV (depending on the particular choices of the two parameters 
that determine the geometry of the space-time, $k$ and $M_5$), a huge enhancement of the cross-section is then possible with respect to standard linearized General Relativity. 

Once fixed the setup we have computed the relevant contributions to the thermally-averaged DM annihilation cross-section
$\langle \sigma \, v \rangle $, taking into accont both virtual KK-graviton and radion/KK-dilaton exchange
as well as  the direct production of radion/KK-dilatons and KK-gravitons. We have then scanned the parameter space of the model 
(represented by $m_{\rm DM}$, $k$ and $M_5$), looking for regions in which the observed relic abundance can be achieved,
$\langle \sigma \, v \rangle \sim \langle \sigma_{\rm FO} \, v \rangle $. 
This region has been  compared with experimental bounds from resonant searches at the LHC Run II and from 
direct and indirect DM detection searches, finding which portion of the allowed parameter space is excluded by data. 
Eventually, we have studied the theoretical unitarity bounds on the mass of the DM particle and on the validity of the CW/LD model
as a consistent low-energy effective theory. We have found that the correct relic abundance may be achieved
in a significant region of the parameter space, corresponding typically to a DM mass of a few TeV's. 

Depending on the spin and the mass of the DM particle, $\langle \sigma_{\rm FO} \, v \rangle $ is reached either through
virtual exchange or direct production of radion/KK-dilatons and/or KK-gravitons. 
For scalar DM particles, we have found that $\langle \sigma_{\rm FO} \, v \rangle $ can be obtained for DM masses in the range 
$m_{\rm DM} \in [1,15]$ TeV and $k \lesssim 6$ TeV. In this case the radion/KK-dilaton virtual exchange increases the 
cross-section for low DM masses (below 1 TeV), thus making possible to achieve $\langle \sigma_{\rm FO} \, v \rangle $
in a much larger portion of the parameter space with respect to the KK-gravitons only case. However, most of this extra region
corresponds to values of $m_{G_1}$ larger than $M_5$ and, thus, in a part of the parameter space where the effective theory
is untrustable. As a consequence, we find no difference between the unstabilized case (no radion/KK-dilatons) and the 
stabilized case in the rigid limit (with radion/KK-dilatons). 
For fermion DM particles the allowed mass range is 
somewhat smaller, $m_{\rm DM} \in [4,15]$ TeV and $k \lesssim 4$ TeV. Eventually, for vector DM particles, the allowed mass range is
somewhat larger, $m_{\rm DM} \in [0.6,15]$ TeV and $k \lesssim 20$ TeV. Notice that the upper limit on the DM mass comes from 
theoretical unitarity bounds. 

Our results for DM in the CW/LD scenario are very similar to those we have found with {\em AdS}$_5$ metric 
(the so-called Randall-Sundrum model) in Ref.~\cite{Folgado:2019sgz}, where we studied only the case of scalar
DM. In the Randall-Sundrum scenario it was known
that, for scalar DM and SM particles localized in the IR brane,  it is not possible to achieve $\langle \sigma_{\rm FO} \, v \rangle $ through the virtual KK-graviton or radion
exchange channel (see also Refs.~\cite{Lee:2013bua,Rueter:2017nbk}). However, we showed that when the DM mass is large enough so that the
direct production of KK-gravitons or radions becomes possible, then the correct relic abundance can be achieved 
for DM particle masses of a few TeV's,  much as in the case of the CW/LD model studied here.
 Notice that the value of $M_5$ needed to achieve
the correct relic abundance in the CW/LD model is $M_5 \in [10,100]$ TeV, whereas
in the Randall-Sundrum scenario the effective coupling $\Lambda$ needed to achieve the freeze-out was in 
$\Lambda \in [10,1000]$ TeV range. In both cases, some hierarchy between the fundamental gravitational scale
(either $M_5$ or $\Lambda$) and the electro-weak scale $\Lambda_{\rm EW}$ is needed.

It is worth to emphasize that in both extra-dimensional scenarios, Randall-Sundrum and CW/LD, it is possible to obtain the correct relic abundance via thermal freeze-out with DM masses in the TeV scale,  so they are already quite constrained by LHC data.
Moreover, most part of the still allowed parameter space may be tested by the LHC Run III 
and by the proposed High-Luminosity LHC. While the prospects for the Randall-Sundrum were already analysed in Ref.~\cite{Folgado:2019sgz}, it would be very interesting to explore in detail the limits that these next LHC phases could set on the CW/LD model.

 \section*{Acknowledgements} 

We thank Matthew McCullough,  Hyun Min Lee and Ver\'onica Sanz for illuminating discussions.
This work has been partially supported by the European Union projects H2020-MSCA-RISE-2015 and H2020-MSCA- ITN-2015//674896-ELUSIVES,
by the Spanish MINECO under grants  FPA2017-85985-P and  SEV-2014-0398, and by Generalitat Valenciana through the ``plan GenT" program (CIDEGENT/2018/019)
and grant PROMETEO/2019/083.

\appendix

\section{Feynman rules}
\label{app:feynman}

We remind in this Appendix the different Feynman rules corresponding to the couplings of DM particles and of SM particles of any spin with KK-gravitons and 
radion/KK-dilatons. 

\subsection{Graviton Feynman rules}
\label{app:gravFR}

The vertex that involves one KK-graviton and two scalars $S$ of mass $m_S$ is given by:
\be
\qquad
\raisebox{-15mm}{\includegraphics[keepaspectratio = true, scale = 1] {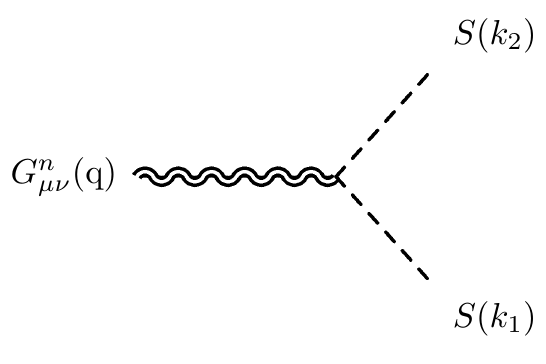}}
\begin {aligned}
=-\frac{i}{\Lambda_n} \left ( m^2_S \eta_{\mu \nu} - C_{\mu \nu \rho \sigma} k_1^{\rho} k_2^{\sigma} \right ) \, ,
\end {aligned}
\ee
where
\be
C_{\mu \nu \alpha \beta} \equiv \eta_{\mu \alpha} \eta_{\nu \beta} + \eta_{\nu \alpha} \eta_{\mu \beta} - \eta_{\mu \nu} \eta_{\alpha \beta} \, .
\ee
This expression can be used for the coupling of both scalar DM and the SM Higgs boson to gravitons.

The vertex that involves one KK-graviton and two fermions $\psi$ of mass $m_\psi$ is given by: 
\be
\qquad
\raisebox{-15mm}{\includegraphics[keepaspectratio = true, scale = 1] {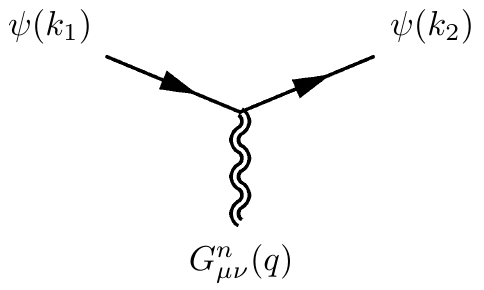}}
\begin {aligned}
=& - \frac{i}{4\Lambda_n}
\left [ \gamma_{\mu} \left ( k_{2 \nu}+k_{1 \nu} \right ) + \gamma_{\nu} \left ( k_{2 \mu}+k_{1 \mu} \right ) \right. \\
&\left. - 2 \eta_{\mu \nu}\left ( \slashed{k_2}+\slashed{k_1}-2m_{\psi} \right )\right ] \, ,
\end {aligned}
\ee
and
\be
\qquad
\raisebox{-15mm}{\includegraphics[keepaspectratio = true, scale = 1] {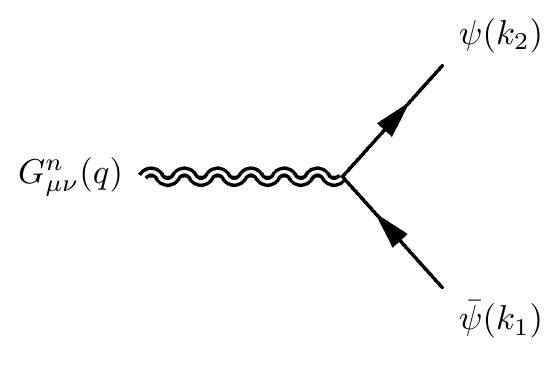}}
\begin {aligned}
=& - \frac{i}{4\Lambda_n}\left [ \gamma_{\mu} \left (k_{2 \nu} - k_{1 \nu} \right ) 
+ \gamma_{\nu} \left (k_{2 \mu} - k_{1 \mu} \right ) \right. \\
&\left. - 2 \eta_{\mu \nu}\left ( \slashed{k_2} -\slashed{k_1} -2m_{\psi} \right )\right ] \, .
\end {aligned}
\ee

The interaction between two vector bosons $V$ of mass $m_V$ and one KK-graviton is given by:
\be
\qquad
\raisebox{-15mm}{\includegraphics[keepaspectratio = true, scale = 1] {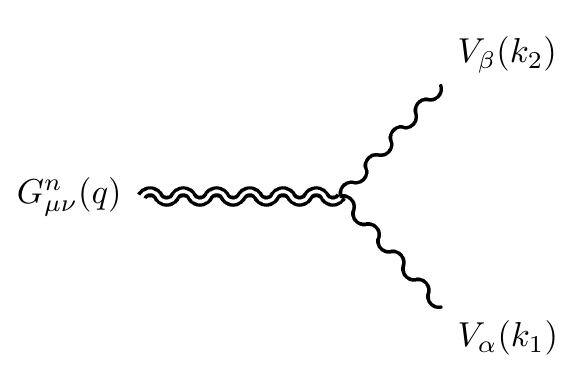}}
\begin {aligned}
=-\frac{i}{\Lambda_n} \left ( m^2_V C_{\mu \nu \alpha \beta} + W_{\mu \nu \alpha \beta} \right ) \, ,
\end {aligned}
\ee
where
\be
W_{\mu \nu \alpha \beta} \equiv B_{\mu \nu \alpha \beta} + B_{\nu \mu \alpha \beta}
\ee
and
\bea
B_{\mu \nu \alpha \beta} &\equiv& \eta_{\alpha \beta}k_{1 \mu}k_{2 \nu} + \eta_{\mu \nu}(k_1 \cdot k_2 \eta_{\alpha \beta} - k_{1 \beta} k_{2 \nu}) \nonumber \\
&-& \eta_{\mu \beta} k_{1 \nu} k_{2 \alpha} + \frac{1}{2}\eta_{\mu \nu}(k_{1 \beta} k_{2 \alpha} - k_1 \cdot k_2 \eta_{\alpha \beta}) \, .
\eea

Eventually, the interaction between two particles ($S, \psi$ or $V_\mu$ depending on their spin) and two KK-gravitons (coming from a second order expansion 
of the metric $g_{\mu\nu}$ around the Minkowski metric $\eta_{\mu\nu}$) is given by:
\be
\label{vertex:scalar_scalar_graviton_graviton}
\qquad
\raisebox{-15mm}{\includegraphics[keepaspectratio = true, scale = 1] {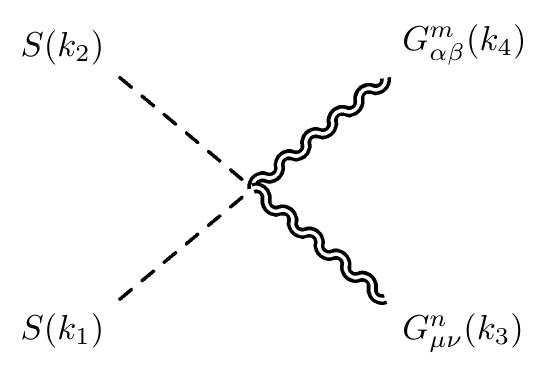}}
\begin {aligned}
=& -\frac{i}{\Lambda_n \Lambda_m} \eta_{\nu \beta} \left ( m^2_S \eta_{\mu \alpha} - C_{\mu \alpha \rho \sigma} k_1^{\rho} k_2^{\sigma} \right ) \, ,
\end {aligned}
\ee
\be
\label{vertex:fermion_fermion_graviton_graviton}
\qquad
\raisebox{-15mm}{\includegraphics[keepaspectratio = true, scale = 1] {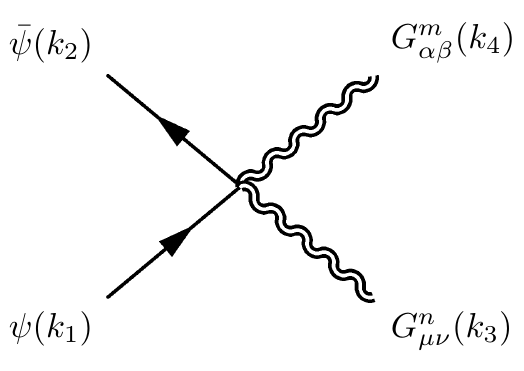}}
\begin {aligned}
=& -\frac{i}{\Lambda_n \Lambda_m} \eta_{\nu \beta} \left [ \gamma_{\mu} \left (k_{1 \alpha} - k_{2 \alpha} \right ) 
+ \gamma_{\alpha} \left (k_{1 \mu} - k_{2 \mu} \right ) \right. \\
&\left. - 2 \eta_{\mu \alpha}\left ( \slashed{k_1} -\slashed{k_2} -2m_\psi \right )\right ] \, ,
\end {aligned}
\ee
\be
\label{vertex:vector_vector_graviton_graviton}
\qquad
\raisebox{-15mm}{\includegraphics[keepaspectratio = true, scale = 1] {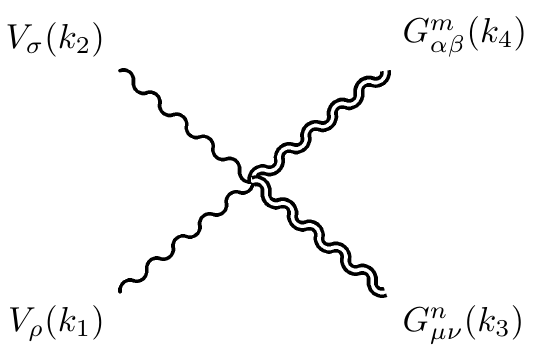}}
\begin {aligned}
=& -\frac{i}{\Lambda_n \Lambda_m} \eta_{\nu \beta} \left ( m^2_V C_{\mu \alpha \rho \sigma} + W_{\mu \alpha \rho \sigma} \right ) \, .
\end {aligned}
\ee

The Feynman rules for the $n=0$ KK-graviton can be obtained by the previous ones by replacing $\Lambda$ with $M_{\rm P}$. 
We do not give here the triple KK-graviton vertex, as it is irrelevant for the phenomenological applications of this paper.

\subsection{Radion/KK-dilaton Feynman rules}
\label{app:radFR}

The radion/KK-dilatons, $\phi_n$, couple with particles localized in the IR-brane with the trace of the energy-momentum tensor, $T = g^{\mu\nu} T_{\mu\nu}$
(in the rigid limit with $\xi = 0$, see Sect.~\ref{sec:rad}). 
The only exception are photons and gluons that, being massless, do not contribute to $T$ at tree-level. 
However, effective couplings of these fields to the radion/KK-dilatons are generated  through quarks and $W$ loops, and the trace anomaly.

The interaction between one radion/KK-dilaton and two scalar fields $S$ of mass $m_S$ is given by:
\be
\qquad
\raisebox{-15mm}{\includegraphics[keepaspectratio = true, scale = 1] {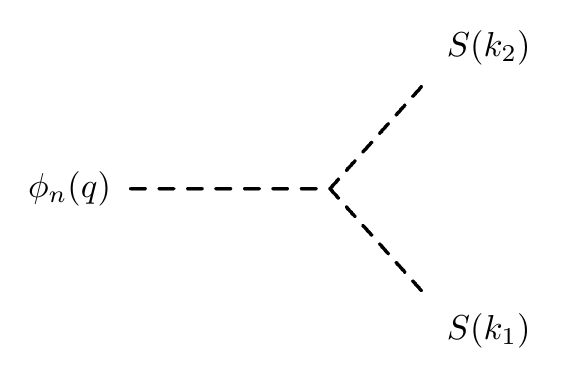}}
\begin {aligned}
=-\frac{2i}{\Lambda_n} \left ( 2m^2_S + k_{1\mu} k_2^\mu \right ) \, .
\end {aligned}
\ee

The vertex that involves one radion/KK-dilaton and two Dirac fermions $\psi$ of mass $m_\psi$ takes the form:
\be
\qquad
\raisebox{-15mm}{\includegraphics[keepaspectratio = true, scale = 1] {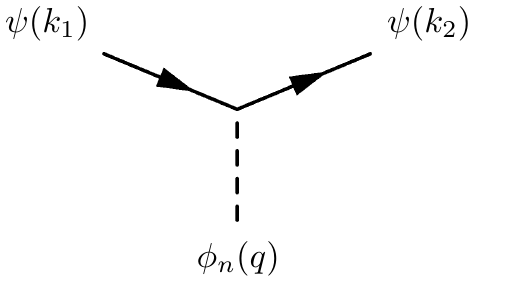}}
\begin {aligned}
=- \frac{i}{2\Lambda_n} \left [8m_{\psi} - 3 \left ( \slashed{k_2}+\slashed{k_1} \right ) \right]  
\end {aligned}
\ee
and:
\be
\qquad
\raisebox{-15mm}{\includegraphics[keepaspectratio = true, scale = 1] {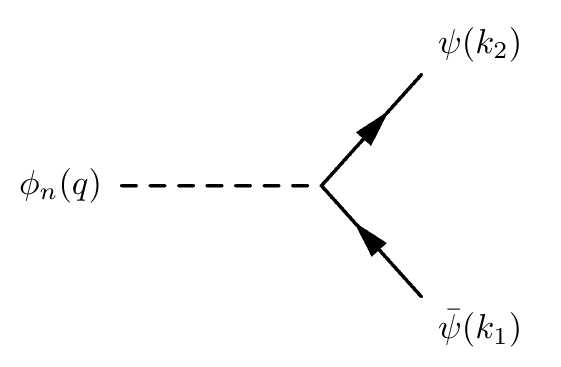}}
\begin {aligned}
=- \frac{i}{2\Lambda_n} \left [ 8m_{\psi} - 3 \left ( \slashed{k_2}-\slashed{k_1} \right ) \right] \, .
\end {aligned}
\ee

The interaction between two massive vector bosons $V$ of mass $m_V$ and one radion/KK-dilaton is given by: 
\be
\qquad
\raisebox{-15mm}{\includegraphics[keepaspectratio = true, scale = 1] {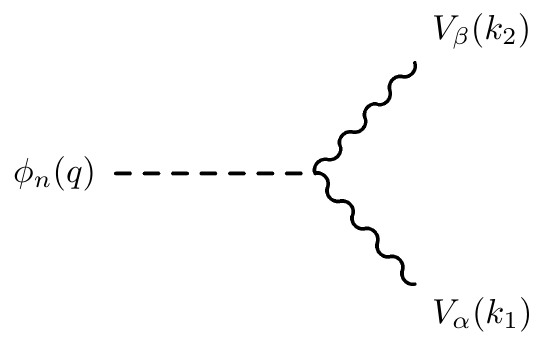}}
\begin {aligned}
=\frac{2i}{\Lambda_n} m^2_V \eta_{\alpha \beta} \, ,
\end {aligned}
\ee
whereas the vertex corresponding to the interaction between two massless SM gauge bosons and one radion/KK-dilaton is: 
\be
\qquad
\raisebox{-15mm}{\includegraphics[keepaspectratio = true, scale = 1] {radion_vector_vector.pdf}}
\begin {aligned}
\label{eq:radiontomasslessvertex}
=\frac{4i\alpha_{i}C_{i}}{8\pi\Lambda_n} \left [ \eta_{\mu\nu}(k_1 \cdot k_2) - k_{1\nu}k_{2\mu}  \right ] \, ,
\end {aligned}
\ee
where $\alpha_{i}=\alpha_{EM}, \alpha_s$ for the case of the photons or gluons, respectively, and \cite{Blum:2014jca}: 
\be
\left \{
\begin{array}{lll}
C_3 &=& b_{IR}^{(3)} - b_{UV}^{(3)} + \frac{1}{2}\sum_q F_{1/2}(x_q) \, , \\
&&\\
C_{EM} &=& b_{IR}^{(EM)} - b_{UV}^{(EM)} + F_1(x_W)  - \sum_q N_cQ_{q}^2F_{1/2}(x_q) \, ,
\end{array}
\right .
\ee
with $x_q = 4m_q/m_r$ and $x_W = 4m_w/m_r$. The values of the one-loop $\beta$-function coefficients $b$ are 
$b_{IR}^{(EM)} - b_{UV}^{(EM)} = 11/3$ and $b_{IR}^{(3)} - b_{UV}^{(3)} = -11 + 2n/3$, 
where $n$ is the number of quarks whose mass is smaller than $m_r/2$.
The explicit form of $F_{1/2}$  and $F_1$ is given by:
\be
\left \{
\begin{array}{lll}
F_{1/2}(x) = 2x[1 + (1-x)f(x)] , \\
&&\\
F_{1}(x) = 2 + 3x + 3x(2-x)f(x) , 
\end{array}
\right .
\ee
with
\be
f(x) = \left \{
\begin{array}{lll}
[\arcsin(1/\sqrt{x})]^2 \hphantom{} \hphantom{} \hphantom{} \hphantom{} \hphantom{} &x>1 , \\
&&\\
-\frac{1}{4}\left[\log\left( \frac{1 + \sqrt{x-1}}{1 - \sqrt{x-1}} \right) - i\pi \right]^2 &x<1 .
\end{array}
\right .
\ee

Eventually, the 4-legs diagrams are given by:
\be
\label{eq:4pointsSSrr}
\qquad
\raisebox{-15mm}{\includegraphics[keepaspectratio = true, scale = 1] {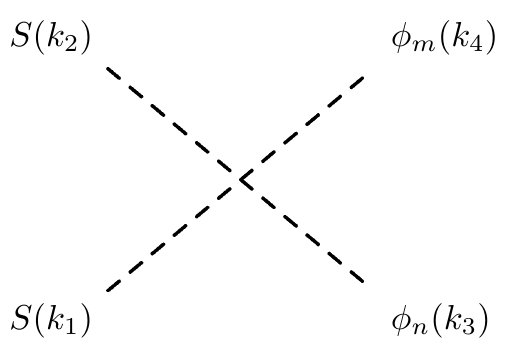}}
\begin {aligned}
=- \frac{i}{3\Lambda^2} \left( 6m_{S}^2 + k_{1\mu} k_2^\mu \right)  \, ,
\end {aligned}
\ee
\be
\label{eq:4pointschichirr}
\qquad
\raisebox{-15mm}{\includegraphics[keepaspectratio = true, scale = 1] {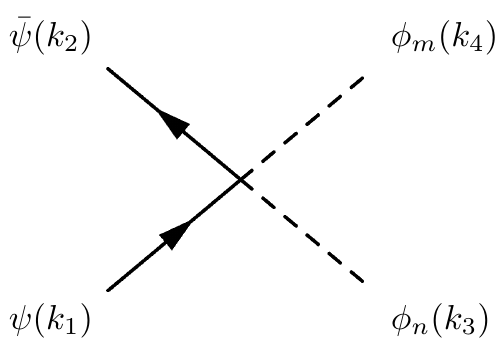}}
\begin {aligned}
= - \frac{i}{2\Lambda_n^2} \left [ 8m_\psi - 3 \left ( \slashed{k_2}-\slashed{k_1} \right ) \right]
\end {aligned}
\ee
and
\be
\label{eq:4pointsVVrr}
\qquad
\raisebox{-15mm}{\includegraphics[keepaspectratio = true, scale = 1] {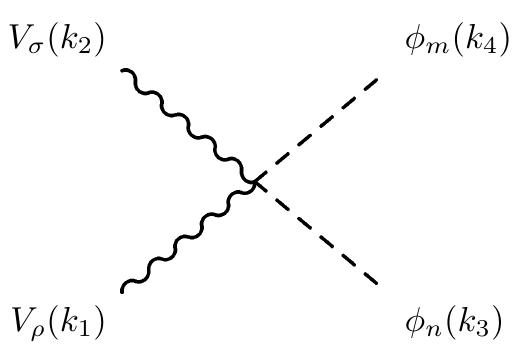}}
\begin {aligned}
= - \frac{2i}{\Lambda_n^2} m^2_V \eta_{\alpha \beta}  \, .
\end {aligned}
\ee

\section{Decay widths}
\label{app:decay}

In this Appendix we compute the decay widths of KK-gravitons and radion/KK-dilatons, using the Feynman rules given in App.~\ref{app:feynman}. 

\subsection{KK-gravitons decay widths}
\label{app:gravdecay}

The KK-graviton can decay into scalar particles (including the Higgs boson, a scalar DM particle and radion/KK-dilatons), fermions (either SM or a fermion DM particle), 
vector bosons (either massive or massless SM gauge bosons or a vector DM particle) and lighter KK-gravitons.

Decay widths of KK-gravitons into SM particles, $\Gamma (G_n \to {\rm SM} \, {\rm SM})$, are all proportional to $1/\Lambda_n^2$. 
In particular, the decay width into SM Higgs bosons  is given by:
\be
\Gamma(G_n \rightarrow HH) = \frac{m_n^3}{960 \, \pi \, \Lambda_n^2}\left(1-\frac{4 \, m_H^2}{m_n^2}\right)^{5/2} \, ,
\ee
where $m_n$ is the mass of the $n$-th KK-graviton (in the main text, this was called $m_{G_n}$, but we prefer here a shorter notation to increase 
readability of the formul\ae).

The decay width of the $n$-th KK-graviton into SM Dirac fermions is given by: 
\be
\Gamma(G_n \rightarrow \bar{\psi} \psi) = \frac{m_n^3}{160 \, \pi \, \Lambda_n^2}\left( 1-\frac{4 \, m_{\psi}^2}{m_n^2} \right)^{3/2}\left( 1+\frac{8 m_{\psi}^2}{3 \, m_n^2} \right).
\ee
The decay width of the $n$-th KK-graviton into two SM massive gauge bosons reads: 
\be
\left \{
\begin{array}{lll}
\Gamma(G_n \rightarrow W^+ W^-) &=& 
\frac{13 \, m_n^3}{480 \, \pi \, \Lambda_n^2} \left( 1-\frac{4 \, m_W^2}{m_n^2}\right)^{1/2}\left(1 + \frac{56 \, m_W^2}{13 \, m_n^2} + \frac{48 \, m_W^4}{13 \, m_n^4} \right) \, , \\
&& \\
\Gamma(G_n \rightarrow ZZ) &=& \frac{13 \, m_n^3}{960 \, \pi \, \Lambda_n^2} \left( 1-\frac{4 \, m_Z^2}{m_n^2}\right)^{1/2}\left( 1 + \frac{56 \, m_Z^2}{13 \, m_n^2} 
+ \frac{48 \, m_Z^4}{13 \, m_n^4} \right)
\, ,
\end{array}
\right .
\ee
whereas the decay width into SM massless gauge bosons is:
\be
\left \{
\begin{array}{lll}
\Gamma(G_n \rightarrow \gamma \gamma) &=& \frac{m_n^3}{80 \, \pi \, \Lambda_n^2} \, ,\\
& & \\
\Gamma(G_n \rightarrow gg) &=& \frac{m_n^3}{10 \, \pi \, \Lambda_n^2} \, .
\end{array}
\right .
\ee
Finally, If $m_n > 2 m_{DM}$, the $n$-th KK-graviton can decay into two DM particles: 
\be
\left \{
\begin{array}{lll}
\Gamma(G_n \rightarrow SS) &=& \frac{m_n^3}{960 \, \pi \, \Lambda_n^2}\left(1-\frac{4 \, m_{\rm DM}^2}{m_n^2}\right)^{5/2} \,  ,\\
& & \\
\Gamma(G_n \rightarrow \bar{\psi} \psi) &=& \frac{m_n^3}{160 \, \pi \,  \Lambda_n^2}\left( 1-\frac{4 \, m_{\rm DM}^2}{m_n^2} \right)^{3/2}\left( 1+\frac{8 \, m_{\rm DM}^2}{3 \, m_n^2} \right)  \,  ,\\
& & \\
\Gamma(G_n \rightarrow VV) &=& \frac{13 \, m_n^3}{960 \, \pi \, \Lambda_n^2} \left( 1-\frac{4 \, m_{\rm DM}^2}{m_n^2}\right)^{1/2}\left( 1 + \frac{56 \, m_{\rm DM}^2}{13 \, m_n^2} 
+ \frac{48 \, m_{\rm DM}^4}{13 \, m_n^4} \right) \, .
\end{array}
\right .
\ee

For completeness, we computed the decay of KK-gravitons into KK-gravitons and radion/KK-dilatons, finding that these contributions are totally negligible. 
For a thorough description of these decays see Ref.~\cite{Giudice:2017fmj}.

\subsection{Radion/KK-dilatons decay widths}
\label{app:radion_decay}

The decay width of the radion/KK-dilatons into SM Higgs boson, is given by:
\be
\Gamma(\phi_n \rightarrow HH) = \frac{m_n^3}{32 \, \pi \, \Lambda_n^2}\left(1 - \frac{4 \, m_H^2}{m_n^2}\right)^{1/2}\left(1+\frac{2 \, m_H^2}{m_n^2}\right)^{2} \, .
\ee
The radion/KK-dilaton decay width into SM Dirac fermions is given by:
\be
\Gamma(\phi_n \rightarrow \bar{\psi} \psi) = \frac{m_n m_{\psi}^2}{8 \, \pi \, \Lambda_n^2}\left( 1-\frac{4 \, m_{\psi}^2}{m_n^2} \right)^{3/2} \, .
\ee
The radion/KK-dilaton decay width into SM massive gauge bosons is: 
\be
\left \{
\begin{array}{lll}
\Gamma(\phi_n \rightarrow W^+W^-) &=& \frac{3 \, m_n^3}{4 \,  \pi  \, \Lambda^2} \left(1 - \frac{4 \, m_W^2}{m_n^2}\right)^{1/2} \left(1 - \frac{m_W^2}{3 \, m_n^2} 
+ \frac{m_W^4}{12 \, m_n^4}\right) \, ,
 \\
&&\\
\Gamma(\phi_n \rightarrow ZZ) &=& \frac{3 \, m_n^3}{8 \,  \pi  \, \Lambda^2} \left(1 - \frac{4 \, m_Z^2}{m_n^2}\right)^{1/2} \left(1 - \frac{m_Z^2}{3 \, m_n^2} 
+ \frac{m_Z^4}{12 \, m_n^4}\right) \, ,
\end{array}
\right .
\ee
whereas the decay width into SM massless gauge bosons is: 
\be
\left \{
\begin{array}{lll}
\Gamma(\phi_n \rightarrow \gamma \gamma) &=& \frac{\alpha_{EM} \, C_{EM} \, m_n^3}{1280 \pi\Lambda^2} \, , \\
&&\\
\Gamma(\phi_n \rightarrow g g) &=& \frac{\alpha_{3} \, C_{3} \, m_n^3}{160 \pi\Lambda^2} \, .
\end{array}
\right .
\ee
If $m_n > 2 m_{DM}$, the $n$-th radion/KK-dilaton can decay into two DM particles: 
\be
\left \{
\begin{array}{lll}
\Gamma(\phi_n \rightarrow SS) &=& \frac{m_n^3}{32 \, \pi \, \Lambda_n^2} \left(1-\frac{4 m_{\rm DM}^2}{m_n^2}\right)^{1/2} \left(1 + \frac{2 \, m_{\rm DM}^2}{m_n^2}\right)^{2} \,  ,\\
& & \\
\Gamma(\phi_n \rightarrow \bar{\psi} \psi) &=& \frac{m_n \,  m_{\rm DM}^2}{8 \, \pi \, \Lambda_n^2} \left( 1 - \frac{4 \, m_{\rm DM}^2}{m_n^2} \right)^{3/2}  \,  ,\\
& & \\
\Gamma(\phi_n \rightarrow VV) &=& \frac{3 \, m_n^3}{8 \,  \pi  \, \Lambda^2}  
\left(1 - \frac{4 \, m_{\rm DM}^2}{m_n^2}\right)^{1/2} 
\left(1 - \frac{m_{\rm DM}^2}{3 \, m_n^2} + \frac{m_{\rm DM}^4}{12 \, m_n^4}\right) \, .
\end{array}
\right .
\ee

We computed the decay of KK-dilatons into KK-gravitons and radion/KK-dilatons, finding that these contributions are totally negligible, as in the case of KK-gravitons.

\section{Sums over KK-gravitons and radion/KK-dilatons}
\label{app:kksum}

In this Appendix we remind the procedure to derive approximated sums over virtual KK-modes following Ref.~\cite{Giudice:2017fmj}. In the main text we have mainly
shown plots using this approximation. However, we show here the degree of accuracy of the approximated sum comparing it with exact results.

Consider the sum over virtual KK-modes that arise both in virtual KK-graviton or virtual radion/KK-dilaton exchange cross-sections: 
\be
\label{eq:suma}
S_{KK} = \sum_{n=1}^{\infty} \frac{1}{\Lambda_n^2}\frac{1}{s-m_n^2 + i m_n \Gamma_n} \, ,
\ee
where $m_n$ is the mass of the $n$-th KK-graviton or radion/KK-dilaton and $\Gamma_n$ its corresponding decay width. 
If $s>k^2$, the modulus squared of the sum over KK-modes is very well approximated by the sum over the KK-modes moduli squared, as the decay widths of the
KK-modes computed in App.~\ref{app:decay} are very small:
\be
\label{eq:suma2}
|S_{KK}|^2 \simeq \sum_{n=1}^{\infty} \frac{1}{\Lambda_n^4}\frac{1}{(s-m_n^2)^2+ m_n^2 \Gamma_n^2} 
\equiv  \sum_{n=1}^{\infty} \frac{1}{\Lambda (m_n)^4} \mathcal{F}(m_n) \, ,
\ee
with ${\cal F}(m_n)$ a function that depends on the mass and the decay width of the virtual KK-modes. We also show explicitly that the $n$-dependence of $\Lambda_n$
in eqs.~(\ref{Lambda_graviton}) and (\ref{Lambda_radion}) arises, indeed, through $m_n$.
The mass difference between two nearby KK-modes, for the typical choices of $k$ and $M_5$ considered in the paper, 
is small enough to approximate the sum by an integral in $m$ starting from the mass of the first KK-mode, $m_1$:
\be
|S_{KK}|^2 \approx \int_{m_1}^{\infty} dm \frac{1}{\Lambda(m)^4} \mathcal{F}(m) \, r_c \, \left(1-\frac{k^2}{m^2}\right)^{-1/2} \, .
\ee
Using the narrow-width approximation for $\mathcal{F}(m)$ 
\be
\mathcal{F}(m) \approx \frac{\pi}{\bar m \, \Gamma (\bar m)} \frac{1}{2\, \sqrt{s}} \delta(\bar m-\sqrt{s}) \, ,
\ee
where $\bar m$ corresponds to the mode for which $m_n \sim \sqrt{s}$ (as enforced by the $\delta$-function), eq.~(\ref{eq:suma2}) can be further approximated as:
\be
\label{eq:KKsumapproximation}
|S_{KK}|^2 \approx \frac{\pi r_c}{2} \frac{1}{\Gamma (\sqrt{s}) \Lambda (\sqrt{s})^4 } \, \left [ \frac{1}{s} \left(1-\frac{k^2}{s}\right)^{-1/2} \right ] \, .
\ee

 Eq.~(\ref{eq:KKsumapproximation}) is valid for both,  KK-gravitons and radion/KK-dilatons.
In the case of KK-gravitons, if we replace $\Lambda_n$ with the expression in eq.~(\ref{Lambda_graviton}), we get:
\be
|S^g_{KK}|^2 \approx \frac{1}{2 \, M_5^6 \, \pi \, r_c} \frac{1}{\Gamma_n|_{m_n \sim \sqrt{s}}} \, 
\left [ \frac{1}{s} \left(1 - \frac{k^2}{s}\right)^{3/2} \right ] \, .
\label{eq:sumaaproximada}
\ee
In the case of radion/KK-dilatons, $\Lambda_n$ is given by eq.~(\ref{Lambda_radion}). Then:
\be
|S^r_{KK}|^2 \approx \frac{8}{729 M_5^6 \pi \, r_c} \frac{1}{\Gamma_n|_{m_n \sim \sqrt{s}}} 
\left [ \frac{1}{s} \left (\frac{k^2}{s}\right)^2 \left(1-\frac{k^2}{s}\right)^{3/2} \left(1-\frac{8 k^2}{9 s}\right)^{-2} \right ] \, ,
\ee
Notice that these expressions are equivalent to an average over the KK-modes.

\begin{figure}[htbp]
\centering
\begin{tabular}{cc}
\includegraphics[width=70mm]{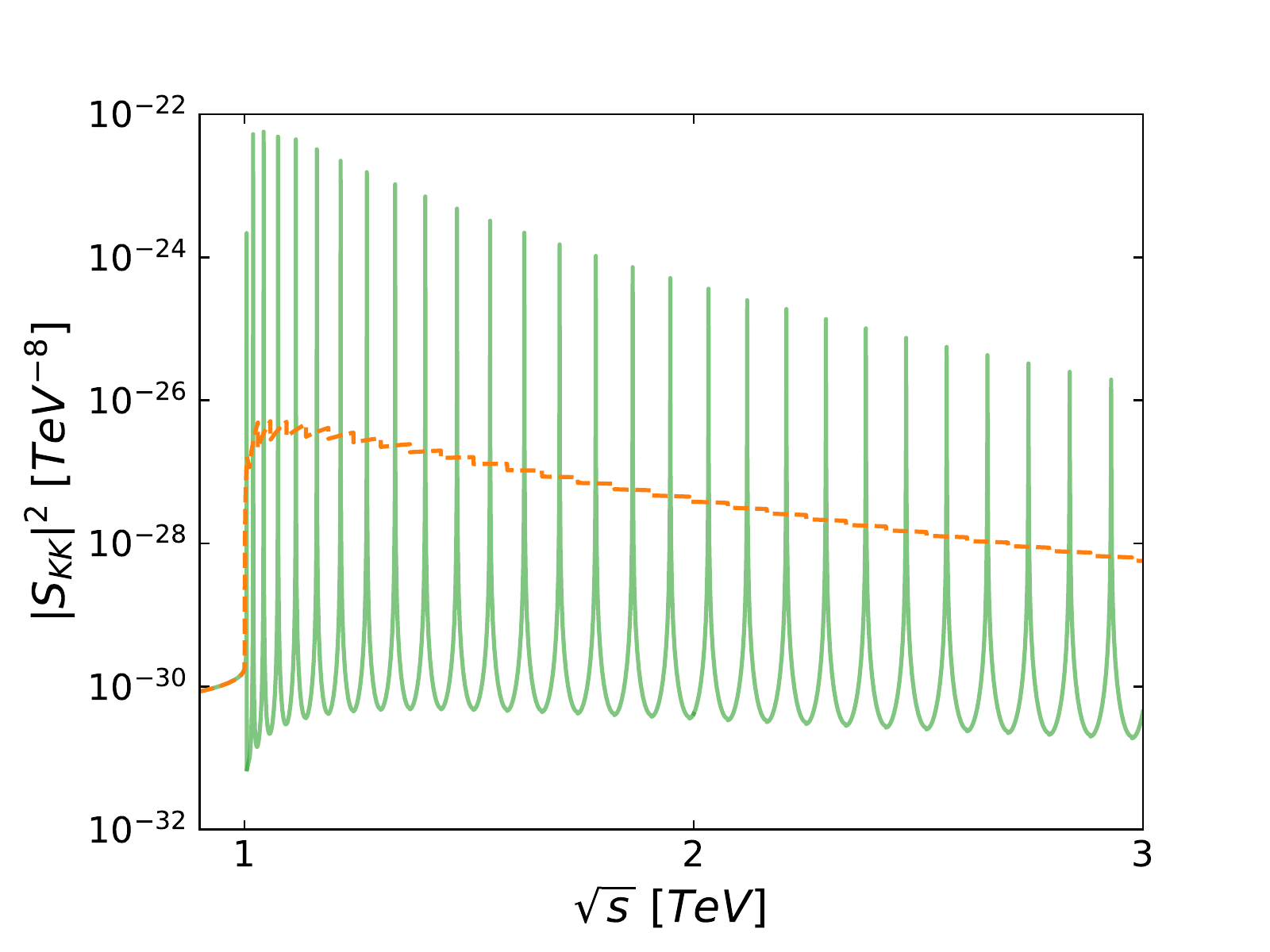} & \includegraphics[width=75mm]{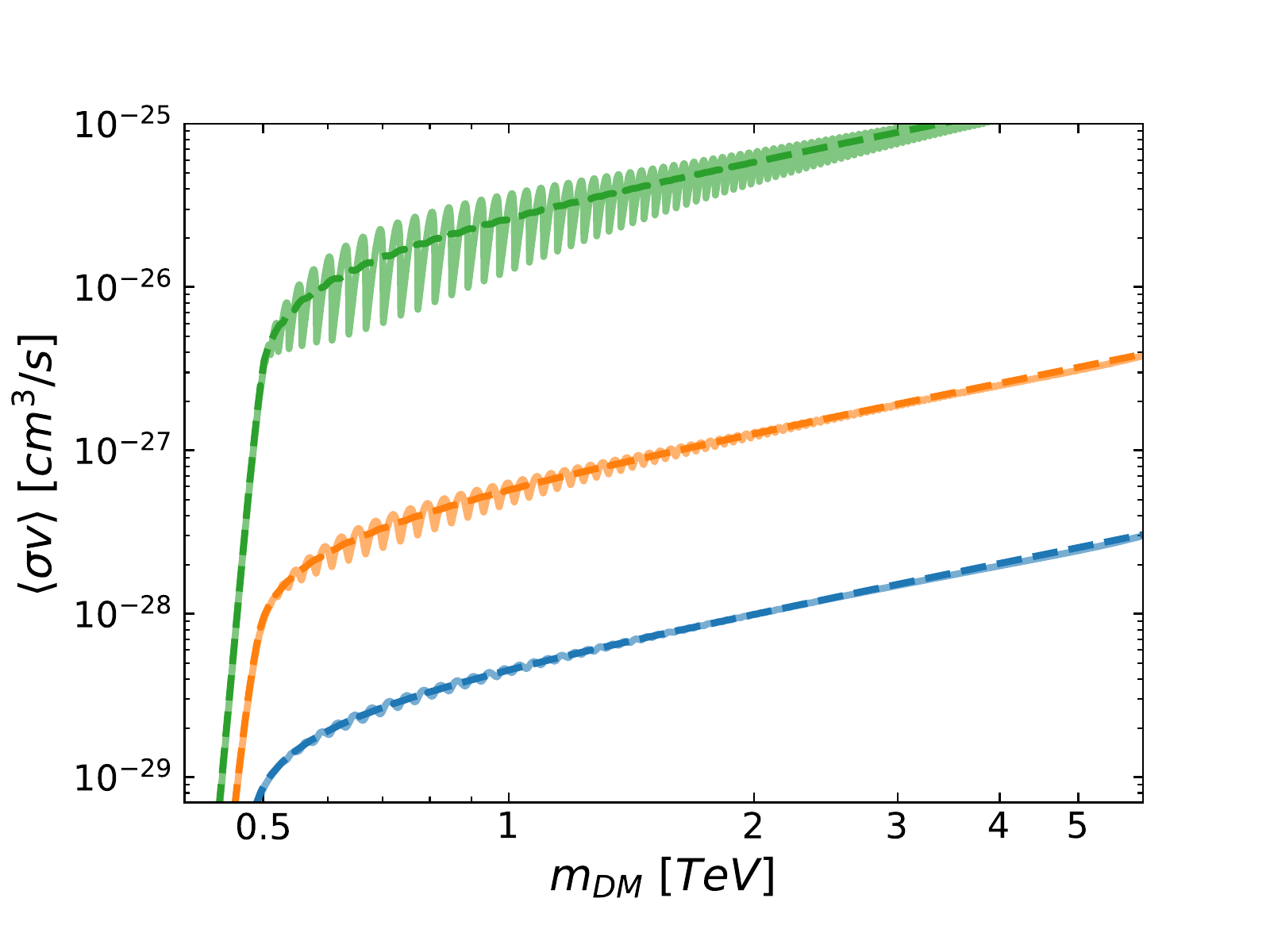}
\end{tabular}
\caption{
Left panel: the sum $|S_{KK}|^2$ for KK-gravitons with $M_5 = 7$ TeV and $k = 1$ TeV. 
The green solid and orange dashed lines represent the result using eq.~(\ref{eq:suma2}) 
and the approximation described in eq.~(\ref{eq:sumaaproximada}), respectively.
Right panel: the thermally-averaged annihilation cross-section through virtual KK-graviton exchange 
for scalar (blue), fermion (orange) an vector (green) DM, with $M_5 = 7$ TeV and $k = 1$ TeV.
Solid lines stand for the exact result, whereas dashed lines represent the approximated one using eq.~(\ref{eq:sumaaproximada}).
}
 \label{fig:KKsum}
\end{figure}

In Fig.~\ref{fig:KKsum} we show 
the comparison between the results for $|S^g_{KK}|^2$ using eqs.~(\ref{eq:suma2}) and (\ref{eq:sumaaproximada}) (left panel),  as well as 
the exact thermally-averaged virtual KK-graviton exchange annihilation cross-section $\langle \sigma v \rangle $ versus the approximated one using 
eq.~(\ref{eq:sumaaproximada}) (right panel), 
for an illustrative choice of $M_5$ and $k$, $M_5 = 7$ TeV and $k = 1$ TeV.
In the left panel we can see how the sum has a very slow onset for $\sqrt{s} \leq k$ summing over the tails of the Breit-Wigner function representing each KK-mode contribution, followed by a very rapidly oscillating behaviour crossing the KK-mode resonances. The difference between being at the dip between two KK-modes or at the peak
can be as large as a factor $10^4$. However, the width of each KK-mode resonance is extremely small and, thus, when summing over many KK-modes the approximated
sum reproduces correctly the collective behaviour of the system.
This is clearly shown in the right panel where we see, that for any spin of 
the DM particle, the exact and approximated sum within the virtual KK-graviton exchange thermally-averaged annihilation cross-section give the same result. 

\section{Annihilation DM Cross section}
\label{app:annihil}

In all the expressions of this Appendix we made use of the so-called {\em velocity expansion} for the DM particles:
\be
s \approx m_{\rm DM}^2(4 + v^2) \, ,
\ee
where $v$ is the relative velocity of the two DM particles.
Within this approximation, the different scalar products for processes in which two DM particles annihilate into two particles (either SM particles, KK-gravitons or radion/KK-dilatons), 
with incoming and outcoming momenta ${\rm DM}(k_1) \, {\rm DM}(k_2) \to {\rm Out}(k_3) \, {\rm Out}(k_4)$, become:
\be
\left \{
\begin{array}{lll}
k_1 \cdot k_4 = k_2 \cdot k_3 \approx 
m_{\rm DM}^2 + \frac{1}{2}m_{\rm DM}^2\sqrt{1-\frac{m_{\rm Out}^2}{m_{\rm DM}^2}}\, \cos{\theta} \, v + \frac{1}{4} m_{\rm DM}^2 \, v^2 \, , \\
&&\\
k_1 \cdot k_3 = k_2 \cdot k_4 \approx 
m_{\rm DM}^2 - \frac{1}{2}m_{\rm DM}^2\sqrt{1-\frac{m_{\rm Out}^2}{m_{\rm DM}^2}}\, \cos{\theta} \, v + \frac{1}{4} m_{\rm DM}^2 \, v^2 \, ,
\end{array}
\right .
\ee
where
\be
\left \{
\begin{array}{lll}
k_1 \cdot k_1 &=& k_2 \cdot k_2 = m_{\rm DM}^2 \, , \\
&& \\
k_3 \cdot k_3 &=& k_4 \cdot k_4 = m_{\rm Out}^2 \, .
\end{array}
\right .
\ee

\subsection{Annihilation through and into KK-gravitons}
\label{app:annihiG}

In the following sections we show the DM annihilation cross-sections through and into KK-gravitons. 
In all of this expressions  $S^g_{KK}$ is the sum over the KK-gravitons given in App.~\ref{app:kksum}. 

\subsubsection{Scalar DM}
\label{app:scalarannihil}

First we start with the scalar Dark Matter. The annihilation cross-section into two SM Higgs bosons is:
\be
\sigma_g(S \, S \rightarrow H \, H) \approx v^3 \, |S^g_{KK}|^2 \frac{m_{\rm DM}^6}{720 \pi} \left(1-\frac{m_H^2}{m_{\rm DM}^2}\right)^{5/2} 
\ee
The annihilation cross-section into two SM massive gauge bosons is:
\be
\left \{
\begin{array}{lll}
\sigma_g(S \, S \rightarrow W^+ \, W^-) &\approx& v^3 \, |S^g_{KK}|^2 \frac{13 \, m_{\rm DM}^6}{360 \pi} 
\left( 1-\frac{m_W^2}{m_{\rm DM}^2} \right)^{1/2} \left( 1 + \frac{14 m_W^2}{13 m_{\rm DM}^2} + \frac{3 m_W^4}{13 m_{\rm DM}^4} \right) \, , \\
&&\\
\sigma_g(S \, S \rightarrow Z \, Z) &\approx & v^3 \, |S^g_{KK}|^2 \frac{m_{13\, \rm DM}^6}{720 \pi}
\left( 1-\frac{m_Z^2}{m_{\rm DM}^2} \right)^{1/2} \left( 1 + \frac{14 m_Z^2}{13 m_{\rm DM}^2} + \frac{3 m_Z^4}{13 m_{\rm DM}^4} \right)  \, ,
\end{array}
\right .
\ee
whereas for two massless gauge bosons we have:
\be
\left \{
\begin{array}{lll}
\sigma_g(S \, S \rightarrow \gamma \, \gamma) &\approx& v^3 \, |S^g_{KK}|^2 \frac{2 m_{\rm DM}^6}{15 \pi} \, , \\
&&\\
\sigma_g(S \, S \rightarrow g \, g) &\approx& v^3 \,  |S^g_{KK}|^2 \frac{m_{\rm DM}^6}{60 \pi} \, .
\end{array}
\right .
\ee
Eventually, the annihilation cross-section into two SM fermions is:
\be
\sigma_g(S \, S \rightarrow \bar{\psi} \, \psi) \approx v^3 \,  |S^g_{KK}|^2 \frac{m_{\rm DM}^6}{120 \pi}
\left( 1 - \frac{m_{\psi}^2}{m_{\rm DM}^2} \right)^{3/2}\left( 1 + \frac{2 m_{\psi}^2}{3 m_{\rm DM}^2} \right) \, .
\ee

As it was shown in Ref.~\cite{Lee:2013bua}, for DM particle masses larger than the mass of a given KK-graviton mode
DM particles may annihilate into two KK-gravitons. In the small velocity approximation, the related cross-section is:
\bea
\sigma_g ( S \, S \rightarrow G_n \, G_m) & \approx & v^{-1} \, \left ( \frac{A^g_S + B^g_S + C^g_S/4}{18432 \, \pi } \right ) \, 
\left ( \frac{1}{\Lambda_n^2 \, \Lambda_m^2 \, m_{\rm DM}^2 \, m_{\rm n}^4 \, m_{\rm m}^4} \right ) \, \nonumber \\
& \times & \sqrt{\left(1+\frac{m_n^2-m_m^2}{4 m_{\rm DM}^2}\right)^2 -\frac{m_n^2}{m_{\rm DM}^2}} \, , 
\label{eq:graviton_real_scalar_dm_cruzados}
\eea
where the three contributions to the cross-section come from the square of the $t$- and $u$-channels 
amplitudes, the square of the 4-points amplitude from the vertex \ref{vertex:scalar_scalar_graviton_graviton} 
and from the interference between the two classes of amplitudes, respectively:
\begin{equation}
\left \{
\begin{array}{lll}
A^g_S &=& \frac{\left [ m_{\rm m}^4
-2 \, m_{\rm m}^2 \, \left ( 4 \, m_{\rm DM}^2 + m_{\rm n}^2 \right )
+ \left(m_{\rm n}^2 - 4 \, m_{\rm DM}^2 \right)^2 
\right ]^4}{2 \left(4 \, m_{\rm DM}^2 - m_{\rm n}^2 - m_{\rm m}^2 \right)^2} \, , \\
  && \\
B^g_S &=& \frac{ \left [ 16 \, m_{\rm DM}^4
-8 \, m_{\rm DM}^2 \, \left(m_{\rm n}^2 + m_{\rm m}^2 \right) + 
\left(m_{\rm n}^2 - m_{\rm m}^2 \right)^2 \right ]^2}{4 \, m_{\rm DM}^2 - m_{\rm n}^2 - m_{\rm m}^2} \,
   \left[16 \, m_{\rm DM}^4
   \left(m_{\rm n}^2 + m_{\rm m}^2 \right) \right.  \\
   &-& \left. 8 \, m_{\rm DM}^2 \, \left( - m_{\rm n}^2 \,
   m_{\rm m}^2 + m_{\rm n}^4 + m_{\rm m}^4 \right) +\left( m_{\rm n}^2 - m_{\rm m}^2\right)^2 \,
   \left(m_{\rm n}^2 + m_{\rm m}^2 \right) \right]  \, , \\
  && \\
C^g_S &=& 256 \, m_{\rm DM}^8 \, \left(13 \, m_{\rm n}^2 \,  m_{\rm m}^2 + 2 \, m_{\rm n}^4 + 2 \, m_{\rm m}^4 \right) 
- 512 \, m_{\rm DM}^6 \, \left(m_{\rm n}^6 + m_{\rm m}^6 \right)  \\   
&&\\
   &+& 32 \, m_{\rm DM}^4 \left (-17 \, m_{\rm n}^6 \, m_{\rm m}^2
   + 98 \,  m_{\rm n}^4 \, m_{\rm m}^4 - 17 \, m_{\rm n}^2 \, m_{\rm m}^6 + 6 \, m_{\rm n}^8 + 6 \, m_{\rm m}^8 \right) \\
&&\\
   &-& 32 \, m_{\rm DM}^2  \left(m_{\rm n}^2 - m_{\rm m}^2 \right)^2 \,
   \left(m_{\rm n}^6 + m_{\rm m}^6 \right) \\
   &&\\
   &+& \left(m_{\rm n}^2 - m_{\rm m}^2 \right)^4 \, \left(13 \, m_{\rm n}^2 \, 
   m_{\rm m}^2 + 2 \, m_{\rm n}^4 + 2 \, m_{\rm m}^4 \right) \, . 
\end{array}
\right .
\end{equation}

In the particular case in which the two KK-gravitons have the same KK-number, m = n, eq.~(\ref{eq:graviton_real_scalar_dm_cruzados}) becomes:
\begin{eqnarray}
\sigma_g ( S \, S \rightarrow G_n \, G_n) & \approx &v^{-1} \frac{4 m_{\rm DM}^2}{9 \, \pi  \, \Lambda_n^2 \Lambda_m^2} 
\frac{(1-r)^{1/2}}{r^4(2-r)^2}  \,  \\
& \times &
\left ( 1 - 3 \, r + \frac{121}{32} \, r^2 - \frac{65}{32} \, r^3 + \frac{71}{128} \, r^4 - \frac{13}{64} \, r^5 + \frac{19}{256} \, r^6 \right ) \, , \nonumber
\label{graviton_real_scalar_dm}
\end{eqnarray}
where $r \equiv (m_{\rm n }/m_{\rm DM})^2$.

\subsubsection{Fermionic case}
\label{app:fermionicannihil}

If the dark matter is a Dirac fermion ($\chi$) the annihilation into two SM Higgs bosons is:
\be
\sigma_g(\bar{\chi} \, \chi \rightarrow H \, H) \approx v \, |S^g_{KK}|^2 \frac{m_{\rm DM}^6}{144 \pi} \left(1-\frac{m_H^2}{m_{\rm DM}^2}\right)^{5/2} 
\ee
The annihilation cross-section into two SM massive gauge bosons is:
\be
\left \{
\begin{array}{lll}
\sigma_g(\bar{\chi} \, \chi \rightarrow W^+ \, W^-) &\approx& v \, |S^g_{KK}|^2 \frac{13 m_{\rm DM}^6}{72 \pi}
\left( 1-\frac{m_W^2}{m_{\rm DM}^2} \right)^{1/2} \left( 1 + \frac{14 m_W^2}{13 m_{\rm DM}^2} + \frac{3 m_W^4}{13 m_{\rm DM}^4} \right) \, , \\
&&\\
\sigma_g(\bar{\chi} \, \chi \rightarrow Z \, Z) &\approx & v \, |S^g_{KK}|^2 \frac{13 m_{\rm DM}^6}{144 \pi}\left( 1-\frac{m_Z^2}{m_{\rm DM}^2} \right)^{1/2} 
\left( 1 + \frac{14 m_Z^2}{13 m_{\rm DM}^2} + \frac{3 m_Z^4}{13 m_{\rm DM}^4} \right)  \, ,
\end{array}
\right .
\ee
whereas for two massless gauge bosons we have:
\be
\left \{
\begin{array}{lll}
\sigma_g(\bar{\chi} \, \chi \rightarrow \gamma \, \gamma) &\approx& v \, |S^g_{KK}|^2 \frac{m_{\rm DM}^6}{12 \pi} \, , \\
&&\\
\sigma_g(\bar{\chi} \, \chi \rightarrow g \, g) &\approx& v \, |S^g_{KK}|^2 \frac{2 \, m_{\rm DM}^6}{3 \pi} \, .
\end{array}
\right .
\ee
Eventually, the annihilation cross-section into two SM fermions is:
\be
\sigma_g(\bar{\chi} \, \chi \rightarrow \bar{\psi} \, \psi) \approx v \, |S^g_{KK}|^2 \frac{m_{\rm DM}^6}{24 \pi}
\left( 1 - \frac{m_{\psi}^2}{m_{\rm DM}^2} \right)^{3/2}\left( 1 + \frac{2m_{\psi}^2}{3 m_{\rm DM}^2} \right) \, .
\ee

As in the case of scalar DM if the $m_{DM}>m_{G_1}$ the $\bar{\psi} \, \psi \rightarrow G_n \, G_m$ channel is open:
\bea
   \sigma_g (\bar{\chi} \, \chi \rightarrow G_n \, G_m) & \approx &v^{-1} \, \left ( \frac{A^g_\chi}{16384 \, \pi} \right ) 
   \left ( \frac{1}{\Lambda_n ^2\Lambda_m ^2 m_{\rm DM}^2 m_{n}^2 m_{m}^2} \right)  \nonumber \\
 & \times &  \sqrt{\left(1+ \frac{m_{n}^2-m_{m}^2}{4 m_{\rm DM}^2} \right)^2 - \frac{m_{n}^2}{m_{\rm DM}^2}} \, . 
\label{eq:graviton_real_fermion_dm_cruzados}
\eea
Notice that, differently from the scalar and vector case, the contribution of the 4-points diagram from the vertex \ref{vertex:fermion_fermion_graviton_graviton} vanishes 
($B^g_\chi = C^g_\chi = 0$). The $t$- and $u$-channel contributions give, instead:
\bea
A^g_\chi = \frac{\left( (m_n^2 - 4m_{\text{DM}}^2)^2 - 2m_m^2( 4m_{\text{DM}}^2 + m_n^2) + m_m^4 \right)^3}{(m_n^2 + m_m^2 - 4m_{\text{DM}}^2)^2}
\eea
In the particular case when two KK-gravitons have the same KK-number, m = n,
 eq.~(\ref{eq:graviton_real_fermion_dm_cruzados}) becomes:
\begin{eqnarray}
\sigma_g (\bar{\chi} \, \chi \rightarrow G_n \, G_n) & \approx &v^{-1} \frac{m_{\rm DM}^2}{16 \, \pi  \, \Lambda_n^4} 
\frac{(1-r)^{7/2}}{r^2(2-r)^2} \, ,
\label{graviton_real_fermion_dm}
\end{eqnarray}
where\footnote{We have found a misprint in Ref.~\cite{Lee:2013bua}: 
the cross-section of fermion DM annihilation into two KK-gravitons scales with $r^{-2}$ as in eq.~(\ref{graviton_real_fermion_dm}), and 
not as $r^{-4}$, as reported in Ref.~\cite{Lee:2013bua}. This is relevant when comparing
results for scalar and vector DM with respect to those for fermion DM as a function of the DM mass (see Sect.~\ref{sec:annihilres}).} 
$r \equiv (m_{\rm n }/m_{\rm DM})^2$.

\subsubsection{Vectorial case}
\label{app:vectorialannihil}

If the dark matter is a spin-1 particle ($X$) the annihilation into two Higgs bosons is:
\be
\sigma_g(X \, X \rightarrow H \, H) \approx v^{-1} \, |S^g_{KK}|^2 \frac{2 m_{\rm DM}^6}{27 \pi} \left(1-\frac{m_H^2}{m_{\rm DM}^2}\right)^{5/2} 
\ee
The annihilation cross-section into two SM massive gauge bosons is:
\be
\left \{
\begin{array}{lll}
\sigma_g(X \, X \rightarrow W^+ \, W^-) &\approx& v^{-1} \, |S^g_{KK}|^2 \frac{52 m_{\rm DM}^6}{27 \pi}\left( 1-\frac{m_W^2}{m_{\rm DM}^2} \right)^{1/2} 
\left( 1 + \frac{14 m_W^2}{13 m_{\rm DM}^2} + \frac{3 m_W^4}{13 m_{\rm DM}^4} \right) \, , \\
&&\\
\sigma_g(X \, X \rightarrow Z \, Z) &\approx & v^{-1} \, |S^g_{KK}|^2 \frac{26 m_{\rm DM}^6}{27 \pi}
\left( 1-\frac{m_Z^2}{m_{\rm DM}^2} \right)^{1/2} \left( 1 + \frac{14 m_Z^2}{13 m_{\rm DM}^2} + \frac{3 m_Z^4}{13 m_{\rm DM}^4} \right)  \, ,
\end{array}
\right .
\ee
whereas for two massless gauge bosons we have:
\be
\left \{
\begin{array}{lll}
\sigma_g(X \, X \rightarrow \gamma \, \gamma) &\approx& v^{-1} \, |S^g_{KK}|^2 \frac{8 m_{\rm DM}^6}{9 \pi} \, , \\
&&\\
\sigma_g(X \, X \rightarrow g \, g) &\approx& v^{-1} \, |S^g_{KK}|^2 \frac{64 \, m_{\rm DM}^6}{9 \pi} \, .
\end{array}
\right .
\ee
The annihilation cross-section into two SM fermions is:
\be
\sigma_g(X \, X \rightarrow \bar{\psi} \, \psi) \approx v^{-1} \, |S^g_{KK}|^2 \frac{12 m_{\rm DM}^6}{27 \pi}\left( 1 - \frac{m_{\psi}^2}{m_{\rm DM}^2} \right)^{3/2}
\left( 1 + \frac{2m_{\psi}^2}{3 m_{\rm DM}^2} \right) \, .
\ee
Eventually, the annihilation into gravitons will be given by:
\bea
\sigma_g ( X \, X \rightarrow G_n \, G_m) & \approx & v^{-1} \, \left ( \frac{A^g_V + B^g_V + C^g_V/2}{331776 \pi } \right ) \, 
\left ( \frac{1}{\Lambda_n^2 \, \Lambda_m^2 \, m_{\rm DM}^2 \, m_{\rm n}^4 \, m_{\rm m}^4} \right ) \, \nonumber \\
& \times & \sqrt{ \left(1+\frac{m_n^2 - m_m^2}{4 m_{\rm DM}^2} \right)^2 - \frac{m_n^2}{m_{\rm DM}^2} } \, ,  
\label{eq:graviton_real_vector_dm_cruzados}
\eea
where:
\begin{equation}
\left \{
\begin{array}{lll}
A^g_V &=& \frac{1}{{\left(-4 m_{\rm DM}^2+m_n^2+m_m^2\right){}^2}} \left[m_{\rm DM}^{16} + 393216 \left(m_n^2+m_m^2\right) m_{\rm DM}^{14} \right. \\ 
&-& \left. 16384 \left(-353m_n^2 m_m^2+m_n^4+m_m^4\right) m_{\rm DM}^{12}  \right. \\ 
&-& \left. \left(m_n^2+m_m^2\right) \left(19 m_n^2 m_m^2+m_n^4+m_m^4\right) m_{\rm DM}^{10} \right. \\ 
&+& \left. 512 \left(2302 m_n^6 m_m^2+3826 m_n^4 m_m^4+2302 m_n^2m_m^6+205 m_n^8+205 m_m^8\right) m_{\rm DM}^8 \right. \\ 
&-& \left. \left(m_n^2+m_m^2\right)\left(-430 m_n^6 m_m^2-602 m_n^4 m_m^4-430 m_n^2 m_m^6+7m_n^8+7 m_m^8\right) m_{\rm DM}^6 \right. \\ 
&-& \left.  \left(1025 m_n^{10} m_m^2+647 m_n^8m_m^4-5562 m_n^6 m_m^6 \right. \right. \\
&+& \left. \left. 647 m_n^4 m_m^8+1025 m_n^2m_m^{10}+21 m_n^{12}+21 m_m^{12}\right) m_{\rm DM}^4 \right. \\ 
&-& \left. \left(m_n^2-m_m^2\right){}^2 \left(m_n^2+m_m^2\right) \left(-67 m_n^6m_m^2-48 m_n^4 m_m^4-67 m_n^2 m_m^6+7 m_n^8+7m_m^8\right) m_{\rm DM}^2 \right. \\ 
&+& \left. \left(m_n^2-m_m^2\right){}^4 \left(208 m_n^6m_m^2+906 m_n^4 m_m^4+208 m_n^2 m_m^6+51 m_n^8+51m_m^8\right)\right] \, ,\\
  && \\
B^g_V &=& 0 \, , \\
  && \\
C^g_V &=& 32768 m_{\rm DM}^{12} -256 \left(-135 m_m^2 m_n^2+74 m_n^4+74 m_m^4\right) m_{\rm DM}^8 \\
&+& 512 \left(m_n^2+m_m^2\right) \left(-43 m_m^2 m_n^2+17 m_n^4+17
   m_m^4\right) m_{\rm DM}^6 \\
   &-& 32 \left(-13 m_m^6 m_n^2-1166 m_m^4 m_n^4-13 m_m^2 m_n^6+42 m_n^8+42 m_m^8\right) m_{\rm DM}^4 \\
   &+& 32 \left(m_n^2-m_m^2\right)^2 \left(m_n^2+m_m^2\right) \left(5 m_m^2 m_n^2+m_n^4+m_m^4\right) m_{\rm DM}^2 \\
   &+& 3 \left(m_n^2-m_m^2\right)^4
   \left(13 m_m^2 m_n^2+2 m_n^4+2 m_m^4\right) \, .
\end{array}
\right .
\end{equation}

In the particular case in which the two KK-gravitons have the same KK-number, m = n,
 eq.~(\ref{eq:graviton_real_vector_dm_cruzados}) becomes:
\begin{eqnarray}
\sigma_g ( X \, X \rightarrow G_n \, G_n) & \approx &v^{-1} \frac{44 m_{\rm DM}^2}{81 \, \pi  \, \Lambda_n^2 \Lambda_m^2} 
\frac{(1-r)^{1/2}}{r^4(2-r)^2}  \, \\
& \times & \left ( 1 + \frac{12}{11} \, r + \frac{351}{44} \, r^2 - \frac{777}{44} \, r^3  + \frac{1105}{176} \, r^4 + \frac{181}{88} \, r^5 + \frac{17}{88} \, r^6 \right ) \, , \nonumber
\label{graviton_real_vector_dm}
\end{eqnarray}
where $r \equiv (m_{\rm n }/m_{\rm DM})^2$.

\subsection{Annihilation through and into radion/KK-dilatons}
\label{app:annihir}

In the following subsections we discuss the different DM annihilation cross sections through and into radion/KK-dilatons, 
using the approximation for the sums over the radion/KK-dilaton modes described in app.\ref{app:kksum}. 
The sum over the dilaton states will be represented as $S^r_{KK}$.

\subsubsection{Scalar case}
\label{app:scalarannihil_dilaton}

The DM annihilation cross-section into two SM Higgs bosons is: 
\be
\sigma_r (S \, S \rightarrow H \, H) \approx v^{-1} \, |S^r_{KK}|^2 \frac{9 \, m_{\rm DM}^6}{\pi} 
\,  \left(1 - \frac{m_H^2}{m_{\rm DM}^2} \right)^{1/2} \, 
\left(1 + \frac{m_h^2}{2 m_{\rm DM}^2}\right)^2 \, ,
\ee

The cross-section for DM annihilation into SM massive gauge bosons is: 
\be
\left \{
\begin{array}{lll}
\sigma_r (S \, S \rightarrow W^+ \, W^-) &\approx & v^{-1} \, |S^r_{KK}|^2 \, \frac{18 \, m_{\rm DM}^6}{\pi} 
 \,  \left( 1 - \frac{m_W^2}{m_{\rm DM}^2} \right)^{1/2} \, \left( 1 - \frac{m_W^2}{m_{\rm DM}^2} + \frac{3 \, m_W^4}{4 m_{\rm DM}^4} \right ) \, , \\
&& \\
\sigma_r (S \, S \rightarrow Z \, Z) & \approx & v^{-1} \, |S^r_{KK}|^2 \, \frac{9 \, m_{\rm DM}^6}{\pi} 
\,  \left( 1 - \frac{m_Z^2}{m_{\rm DM}^2} \right)^{1/2} \, \left( 1 - \frac{m_Z^2}{m_{\rm DM}^2} + \frac{3 \, m_Z^4}{4 m_{\rm DM}^4} \right) \, . \\
&&
\end{array}
\right .
\ee
The DM annihilation into photons and gluons is proportional to the vertex in eq.~(\ref{eq:radiontomasslessvertex}). 
The corresponding expressions for the cross-sections are:
\be
\left \{
\begin{array}{lll}
\sigma_r (S \, S \rightarrow \gamma \, \gamma) &\approx & v^{-1} \, |S^r_{KK}|^2 \, 
\frac{9 \, m_{\rm DM}^6 \, \alpha_{EM} \, C_{EM}}{8 \, \pi^3} \, , \\
&& \\
\sigma_r (S \, S \rightarrow g \, g) & \approx & v^{-1} \, |S^r_{KK}|^2 \, 
\frac{9 \, m_{\rm DM}^6 \, \alpha_{3} \, C_{3}}{\pi^3} \, .
\end{array}
\right .
\ee
The DM annihilation cross-section into SM fermions is given by:
\be
\sigma_r (S \, S \rightarrow \bar{\psi} \, \psi) \approx v^{-1} \, |S^r_{KK}|^2 \, \frac{9 \, m_{\rm DM}^4 \, m_{\psi}^2}{\pi} \,  \left( 1 - \frac{m_{\psi}^2}{m_{\rm DM}^2} \right)^{3/2}.
\ee
Eventually, the DM annihilation cross-section into two radion/KK-dilatons is given by:
\be
\sigma_g ( S \, S \rightarrow \phi_n \, \phi_m) \approx v^{-1} \, \frac{A^r_S+B^r_S+C^r_S}{64 \pi \Lambda^2_n\Lambda_m^2 m_{\rm DM}^2} 
 \times  \sqrt{\left(1+\frac{m_n^2-m_m^2}{4 m_{\rm DM}^2} \right)^2 - \frac{m_n^2}{m_{\rm DM}^2} }
\ee
where, as in the case of KK-gravitons, the three contributions to the cross-section come from the square of the $t$- and $u$-channels 
amplitudes ($A^r_S$), the square of the 4-points amplitude from vertex \ref{eq:4pointsSSrr} ($C^r_S$)
and from the interference between the two classes of diagrams ($B^r_S$), respectively:
\begin{equation}
\left \{
\begin{array}{lll}
A^r_S &=& \frac{\left [64 m_{\rm DM}^2 + (m_n^2 - m_m^2)^2 \right ]^2}{ (-4 m_{\rm DM}^2 + m_n^2 + m_m^2)^2} \, , \\
  && \\
B^r_S &=&  \frac{28 \left [64 m_{\rm DM} + (m_n^2 - m_m^2)^2 \right ]}{(-4 m_{\rm DM}^2 + m_n^2 + m_m^2)}  \, , \\
  && \\
C^r_S &=&  196 \, m_{\rm DM}^4 \, . 
\end{array}
\right .
\end{equation}
where $(m_n, \Lambda_n)$ and $(m_m, \Lambda_m)$ are the masses and coupling of the $n$-th and $m$-th radion/KK-dilatons modes, 
respectively.

\subsubsection{Fermionic case}
\label{app:fermionicannihil_dilaton}

If the Dark Matter is a Dirac fermion ($\chi$) the annihilation into two SM Higgs bosons is:
\be
\sigma_r (\bar{\chi} \, \chi \rightarrow H \, H) \approx v \, |S^r_{KK}|^2 \frac{m_{\rm DM}^6}{8 \, \pi} 
\,  \left(1 - \frac{m_H^2}{m_{\rm DM}^2} \right)^{1/2} \, 
\left(1 + \frac{m_H^2}{2 m_{\rm DM}^2}\right)^2 \, ,
\ee
The annihilation cross-section into two SM massive gauge bosons is:
\be
\left \{
\begin{array}{lll}
\sigma_r (\bar{\chi} \, \chi \rightarrow W^+ \, W^-) &\approx & v \, |S^r_{KK}|^2 \, \frac{m_{\rm DM}^6}{4 \, \pi} 
 \,  \left( 1 - \frac{m_W^2}{m_{\rm DM}^2} \right)^{1/2} \, \left( 1 - \frac{m_W^2}{m_{\rm DM}^2} + \frac{3 \, m_W^4}{4 m_{\rm DM}^4} \right ) \, , \\
&& \\
\sigma_r (\bar{\chi} \, \chi  \rightarrow Z \, Z) & \approx & v \, |S^r_{KK}|^2 \, \frac{m_{\rm DM}^6}{8 \, \pi} 
\,  \left( 1 - \frac{m_Z^2}{m_{\rm DM}^2} \right)^{1/2} \, \left( 1 - \frac{ m_Z^2}{m_{\rm DM}^2} + \frac{3 \, m_Z^4}{4 m_{\rm DM}^4} \right) \, . \\
&&
\end{array}
\right .
\ee
whereas for two massless gauge bosons we have:
\be
\left \{
\begin{array}{lll}
\sigma_r (\bar{\chi} \, \chi \rightarrow \gamma \, \gamma) &\approx & v \, |S^r_{KK}|^2 \, 
\frac{m_{\rm DM}^6 \, \alpha_{EM} \, C_{EM}}{16 \, \pi^3} \, , \\
&& \\
\sigma_r (\bar{\chi} \, \chi \rightarrow g \, g) & \approx & v \, |S^r_{KK}|^2 \, 
\frac{m_{\rm DM}^6 \, \alpha_{3} \, C_{3}}{2 \, \pi^3} \, .
\end{array}
\right .
\ee
The DM annihilation cross-section into two SM fermions is:
\be
\sigma_r (\bar{\chi} \, \chi \rightarrow \bar{\psi} \, \psi) \approx v \, |S^r_{KK}|^2 \, \frac{m_{\rm DM}^4 \, m_{\psi}^2}{8 \, \pi} \,  \left( 1 - \frac{m_{\psi}^2}{m_{\rm DM}^2} \right)^{3/2}.
\ee
Eventually, the annihilation directly into dilatons is given by:
\be
\sigma_g (\bar{\chi} \, \chi  \rightarrow \phi_n \, \phi_m) \approx v \, \frac{A^r_\chi+B^r_\chi+C^r_\chi}{13824 m_{DM}^2 \pi \Lambda^2_n\Lambda_m^2} \sqrt{\left(1+ \frac{m_n^2-m_m^2}{4m_{\rm DM}^2}\right)^2 - \frac{m_n^2}{m_{\rm DM}^2} }
\ee
where:
\begin{equation}
\left \{
\begin{array}{lll}
A^r_\chi &=& \frac{m_{DM}^4}{\left(-4 m_{\text{DM}}^2+m_{\text{n}}^2+m_{\text{m}}^2\right){}^4}  \left[4 m_{\text{m}}^6 \left(419 m_{\text{n}}^2-1804 m_{\text{DM}}^2\right) \right. \\
&+& \left. 2 m_{\text{m}}^4 \left(-10312
   m_{\text{DM}}^2 m_{\text{n}}^2+21648 m_{\text{DM}}^4+3273 m_{\text{n}}^4\right) \right. \\
    &-& \left. 4 m_{\text{m}}^2 \left(1804
   m_{\text{DM}}^2-419 m_{\text{n}}^2\right) \left(m_{\text{n}}^2-4 m_{\text{DM}}^2\right){}^2+451 \left(m_{\text{n}}^2-4
   m_{\text{DM}}^2\right){}^4+451 m_{\text{m}}^8\right] \, , \\
  && \\
B^r_\chi &=& 0  \, , \\
  && \\
C^r_\chi &=&  3 m_{\text{DM}}^4 \, . 
\end{array}
\right .
\end{equation}
and where $(m_n, \Lambda_n)$ and $(m_m, \Lambda_m)$ are the masses and coupling of the $n$-th and $m$-th radion/KK-dilatons modes, 
respectively.

\subsubsection{Vectorial case}
\label{app:vectorialannihil}

If the Dark Matter is a spin-1 particle ($X$) the annihilation into two SM Higgs bosons is:
\be
\sigma_r (X \, X \rightarrow H \, H) \approx v^{-1} \, |S^r_{KK}|^2 \frac{m_{\rm DM}^6}{3 \, \pi} 
\,  \left(1 - \frac{m_H^2}{m_{\rm DM}^2} \right)^{1/2} \, 
\left(1 + \frac{m_H^2}{2 m_{\rm DM}^2}\right)^2 \, ,
\ee
The annihilation cross-section into two SM massive gauge bosons is:
\be
\left \{
\begin{array}{lll}
\sigma_r (X \, X \rightarrow W^+ \, W^-) &\approx & v^{-1} \, |S^r_{KK}|^2 \, \frac{4 m_{\rm DM}^2 \, m_W^4}{3 \, \pi} 
 \,  \left( 1 - \frac{m_W^2}{m_{\rm DM}^2} \right)^{1/2} \, \left( 1 - \frac{3 m_W^2}{4 m_{\rm DM}^2} + \frac{ m_W^4}{8 m_{\rm DM}^4} \right ) \, , \\
&& \\
\sigma_r (X \, X  \rightarrow Z \, Z) & \approx & v^{-1} \, |S^r_{KK}|^2 \, \frac{2 m_{\rm DM}^2 \, m_Z^4}{3 \, \pi} 
\,  \left( 1 - \frac{m_Z^2}{m_{\rm DM}^2} \right)^{1/2} \, \left( 1 - \frac{3 m_Z^2}{4 m_{\rm DM}^2} + \frac{m_Z^4}{8 m_{\rm DM}^4} \right) \, . \\
&&
\end{array}
\right .
\ee
whereas for two massless gauge bosons we have:
\be
\left \{
\begin{array}{lll}
\sigma_r (X \, X \rightarrow \gamma \, \gamma) &\approx & v^{-1} \, |S^r_{KK}|^2 \, 
\frac{3 \, m_{\rm DM}^6 \, \alpha_{EM} \, C_{EM}}{8 \, \pi^3} \, , \\
&& \\
\sigma_r (X \, X \rightarrow g \, g) & \approx & v^{-1} \, |S^r_{KK}|^2 \, 
\frac{3 \, m_{\rm DM}^6 \, \alpha_{3} \, C_{3}}{\pi^3} \, .
\end{array}
\right .
\ee
The DM annihilation cross-section into two SM fermions is:
\be
\sigma_r (X \, X \rightarrow \bar{\psi} \, \psi) \approx v^{-1} \, |S^r_{KK}|^2 \, \frac{m_{\rm DM}^4 \, m_{\psi}^2}{3 \, \pi} \,  \left( 1 - \frac{m_{\psi}^2}{m_{\rm DM}^2} \right)^{3/2}.
\ee
Eventually, the annihilation cross-section into two radion/KK-dilatons is given by:
\be
\sigma_g (X \, X \  \rightarrow \phi_n \, \phi_m) \approx v^{-1} \, \frac{A^r_V+B^r_V+C^r_V}{20736 \, \pi \, \Lambda^2_n \, \Lambda_m^2 m_{\rm DM}^2} \sqrt{\left(1 + \frac{m_n^2-m_m^2}{4m_{\rm DM}^2} \right)^2 - \frac{m_n^2}{m_{\rm DM}^2} }
\ee
where:
\begin{equation}
\left \{
\begin{array}{lll}
A^r_V &=& \frac{1}{ (-4 m_{\rm DM}^2 + m_n^2 + m_m^2)^2}\left[-512 \left(m_n^2+m_m^2\right) m_{\rm DM}^6
+128 \left(m_n^4+m_m^4\right) m_{\rm DM}^4 \right. \\
&-& \left. 16 \left(m_n^2-m_m^2\right)^2
   \left(m_n^2+m_m^2\right) m_{\rm DM}^2+\left(m_n^2-m_m^2\right)^4+1536 m_{\rm DM}^8\right] \, , \\
  && \\
B^r_V &=& 0  \, , \\
  && \\
C^r_V &=& 12 m_{\rm DM}^4 \, . 
\end{array}
\right .
\end{equation}
and where $(m_n, \Lambda_n)$ and $(m_m, \Lambda_m)$ are the masses and coupling of the $n$-th and $m$-th radion/KK-dilatons modes, 
respectively.

\subsection{Annihilation into one KK-graviton and one radion/KK-dilaton}

It exists another channel that was not previously considered in the literature: DM annihilation into one KK-graviton 
and one radion/KK-dilaton. The cross-section for this process is given by the following expressions:
\begin{equation}
\left \{
\begin{array}{lll}
   \sigma_{gr} (S \, S \rightarrow G_n \, r_m) & \approx &v^{-1} \, \left ( \frac{A^{gr}_S}{9216 \pi} \right ) 
   \left ( \frac{1}{\Lambda_{g,n} ^2 \, \Lambda_{r,m} ^2 m_{\rm DM}^2 m_{g,n}^4} \right) \,  \frac{1}{\left(-4 m_{\text{DM}}^2+m_{g,n}^2+m_{r,m}^2\right)^2} \\
 & \times &  \sqrt{\left(1+ \frac{m_{g,n}^2-m_{r,m}^2}{4 m_{\rm DM}^2} \right)^2 - \frac{m_{g,n}^2}{m_{\rm DM}^2}} \, , \\
  && \\
   \sigma_{gr}(\bar{\chi} \, \chi \rightarrow G_n \, r_m) & \approx &v^{-1} \, \left ( \frac{A^{gr}_\chi}{576 \pi} \right ) 
   \left ( \frac{1}{\Lambda_{g,n} ^2\, \Lambda_{r,m}^2  m_{g,n}^2} \right) \, \frac{1}{\left(-4 m_{\text{DM}}^2+m_{g,n}^2+m_{r,m}^2\right)^2}    \nonumber \\
 & \times &  \sqrt{\left(1+ \frac{m_{g,n}^2-m_{r,m}^2}{4 m_{\rm DM}^2} \right)^2 - \frac{m_{g,n}^2}{m_{\rm DM}^2}} \, , \\
  && \\
   \sigma_{gr} (V \, V \rightarrow G_n \, r_m) & \approx &v^{-1} \, \left ( \frac{A^{gr}_V}{82944 \pi} \right ) 
   \left ( \frac{1}{\Lambda_{g,n}^2\Lambda_{r,m}^2 m_{\rm DM}^2 m_{g,n}^4} \right) \,  \frac{1}{\left(-4 m_{\text{DM}}^2+m_{g,n}^2+m_{r,m}^2\right)^2}   \\
 & \times &  \sqrt{\left(1+ \frac{m_{g,n}^2-m_{r,m}^2}{4 m_{\rm DM}^2} \right)^2 - \frac{m_{g,n}^2}{m_{\rm DM}^2}} \, ,
\end{array}
\right .
\end{equation}
where the value of $A^{gr}$ is given by:
\begin{equation}
\left \{
\begin{array}{lll}
A^{gr}_S &=& \left(m_{g,n}^2-m_{r,m}^2\right)^2 \left[ -2 m_{r,m}^2 \left(4 m_{\text{DM}}^2+m_{g,n}^2 \right ) +\left(m_{g,n}^2-4
   m_{\text{DM}}^2\right){}^2+m_{r,m}^4\right ]^2 \, , \\
  && \\
A^{gr}_\chi &=& \left(2 m_{\text{DM}}-m_{g,n}-m_{r,m}\right) \left(2 m_{\text{DM}}+m_{g,n}-m_{r,m}\right) \\
 &\times &\left(2 m_{\text{DM}}-m_{g,n}+m_{r,m}\right) \left(2 m_{\text{DM}}+m_{g,n}+m_{r,m}\right) \\
 &\times&  \left [8 m_{\text{DM}}^2 \left(7 m_{g,n}^2-3 m_{r,m}^2\right)+48 m_{\text{DM}}^4 + 3 \left(m_{g,n}^2-m_{r,m}^2\right)^2 \right ] \, , \\
  && \\
A^{gr}_V &=& \, 4096 m_{\text{DM}}^{10} \left(3 m_{g,n}^2-7 m_{r,m}^2\right)+256 m_{\text{DM}}^8 \left(-106 m_{g,n}^2 m_{r,m}^2+93 m_{g,n}^4+53 m_{r,m}^4\right) \\
&+& 256
   m_{\text{DM}}^6 \left(-63 m_{g,n}^4 m_{r,m}^2+57 m_{g,n}^2 m_{r,m}^4+67 m_{g,n}^6-13 m_{r,m}^6\right) \\
   &+& 64 m_{\text{DM}}^4 \left(m_{g,n}^2-m_{r,m}^2\right){}^2
   \left(-34 m_{g,n}^2 m_{r,m}^2+17 m_{g,n}^4+7 m_{r,m}^4\right) \\ 
   &+& 32 m_{\text{DM}}^2 \left(m_{g,n}^2-m_{r,m}^2\right){}^4 \left(4 m_{g,n}^2-m_{r,m}^2\right)+24576
   m_{\text{DM}}^{12}+\left(m_{g,n}^2-m_{r,m}^2\right){}^6 \, . 
\end{array}
\right .
\end{equation}
In all of these expressions we have used $(m_{g,n},\Lambda_{g,n})$ and $(m_{r,m},\Lambda_{r,m})$ for the mass and coupling of the $n$-th KK-graviton and of the $m$-th radion/KK-dilaton, respectively. Notice that for this particular channel it does not exists a four-legs vertex.

\bibliography{bibliografia} 

\end{document}